\documentclass[aps,pra,twocolumn,groupedaddress]{revtex4}
\usepackage{amsfonts}

\usepackage{graphicx}
\usepackage{dcolumn}
\usepackage{bm}

\begin{document}

\title{Optical response of a misaligned and suspended Fabry-Perot cavity}

\author{G. Cella$^1$, A. Di Virgilio$^1$, P. La Penna$^{1,2}$}
\affiliation{$^1$ INFN, sez. Pisa \\
$^2$ EGO European Gravitational Observatory}
\author{V. D'Auria$^1$, A. Porzio$^{2,3}$, I. Ricciardi$^{1,4}$, S. Solimeno$^{1,3,4}$}
\affiliation{$^1$ Dip. Scienze Fisiche, Univ. Federico II \\
$^2$ "Coherentia", CNR--INFM, Napoli\\
$^3$ CNISM, Unit\'a di Napoli \\
$^4$ INFN, sez. Napoli}

\date{\today}

\begin{abstract}
The response to a probe laser beam of a suspended, misaligned and detuned
optical cavity is examined. A five degree of freedom model of the
fluctuations of the longitudinal and transverse mirror coordinates is
presented. Classical and quantum mechanical effects of radiation pressure
are studied with the help of the optical stiffness coefficients and the
signals provided by an FM sideband technique and a quadrant detector, for
generic values of the product $\varpi \tau $ of the fluctuation frequency
times the cavity round trip. A simplified version is presented for the case
of small misalignments. Mechanical stability, mirror position entanglement
and ponderomotive squeezing are accommodated in this model. Numerical plots
refer to cavities under test at the so-called Pisa LF facility.
\end{abstract}

\maketitle

\section{Introduction and notation}

Optical cavities are generally studied by assuming a single mode excitation
and ignoring the photon scattering by mirror reflections into other modes. A
single mode description is no more reliable when a slight misalignment is
sufficient to excite different modes. This situation is met in almost
concentric or plane--parallel or confocal configurations. In other cases the
weak amplitudes of modes falling outside the resonance bandwidth are of
interest. For example, the sensitivity of interferometers with cavities
placed in their arms depends on the contribution of higher modes as well as
the error signals used for longitudinal and angular alignments.

Optical cavities are generally stabilized, in length, by the Drever-Pound
(D-P) technique \cite{Drever} working with odd harmonics of the phase
modulated laser beam.

In order to deal with a large variety of situations the model discussed in
this paper accounts for misalignment, detuning and a generic spectrum of
harmonics. Faced with the possibility of working with approximate
expressions it has been preferred to simplify exact solutions at the end of
the calculation. This option avoids difficulty of making adequate
approximations in presence of a large number of parameters.

This strategy can be useful for the design studies necessary for the
development of future gravitational antennas. It gives the possibility to
investigate noise contributions coming from all optical and mechanical
degrees of freedom. It can be also used for studying instabilities, optical
spring effect, entanglement and radiative pressure squeezing associated to
both axial and angular fluctuations for any degree of detuning, misalignment
and mismatch.

The present work grew up from the study of short and large spot size
resonators \cite{La Penna} implemented at LFF (Low Frequency Facility \cite
{Bernardini}) a facility dedicated to testing new mechanical suspensions,
controls and mirrors for the VIRGO interferometric gravitational antenna,
and studying the effects of radiation pressure, mirror and suspensions
thermal noise.

Main features of the LFF are the use of suspended mirrors and the
possibility of confining large section cavity modes. The mirrors hang from
multipendula which guarantee a drastic reduction of the seismic noise above
the resonance frequencies of the mechanical modes. The phase-modulated light
reflected by the cavity is used by a Pound-Drever apparatus \cite{Drever}
both for stabilizing the cavity length, and measuring the noise spectrum.
Several papers analyzed the dynamic and the alignment of cavities sharing
some of the LFF features \cite{Fritschel,Dorsel,Moss}. Numerous specialized
studies have been produced by research groups of VIRGO, LIGO, TAMA and GEO
projects \cite{web}.

Suspended cavities have been analyzed by several authors in different
contexts, all sharing the common feature of using a system of Langevin
equations for both the mechanical and electromagnetic modes. The coupling of
cavity and mechanical modes is represented by suitable potentials \cite
{Law,Pace} leading to a complex interplay between cavity mode amplitudes,
mirror positions and orientation fluctuations. In this paper the resonator
is regarded as a mechanical Langevin system driven by thermal sources and
shot noise. This is done by-passing the Hamiltonian approach and hiding the
optical modes fluctuations into the mechanical ones by generalizing an
approach introduced in \cite{Deruelle}. So doing, the Langevin system
contains ponderomotive terms, connected with the classical part of the laser
beam and the shot noise. The seismic noise affecting the mechanical system
has been neglected. Once known, its local spectral density can be easily
added to the thermal one.

Thermal motion of mechanical oscillators has been modelled as standard
Brownian motion \cite{Gardiner}, possibly corrected by Diosi for preserving
the quantum mechanical commutation relations \cite{Jacobs}, or by
non-Lindblad master equations (ME) \cite{Munro,Giovannetti}. Accordingly, in
the present model different thermal correlation functions have been
introduced.

The quantum field fluctuations (shot-noise) are accounted for by splitting
each mode amplitude in a classical and a quantum parts \cite{Pace,
Barchielli} and relying on the input-output theory \cite{Collet}.

Radiation pressure can lead to mechanical instabilities, as predicted by
Braginsky and Manukin \cite{Braginsky&Manukin}. Acting on the suspended
mirrors it provides a spring action which either depresses or reinforces any
perturbation \cite{Deruelle,Sidles,spring,Braginsky}. It may also be used to
mechanically entangle the two mirrors \cite{Mancini} or to enhance the
squeezing of the output field \cite{Heidmann}.

\begin{figure}
\includegraphics[width=7cm]{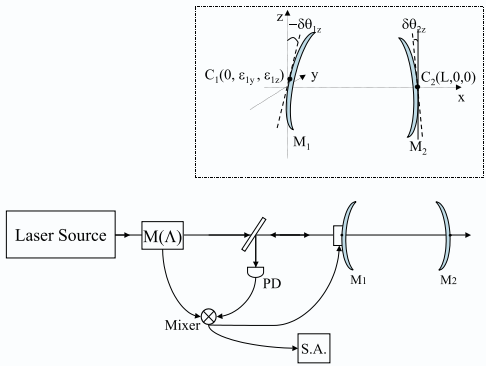}
\caption{
Typical optical layout of the apparatus examined in the present paper.
The laser beam is phase modulated at frequency $\Lambda $ by
the modulator $M\left( \Lambda \right)$. The modulated bema drives the cavity.
The front $M_{1}$ and back $M_{2}$ mirrors are suspended to 
multipendulum chains. The beam reflected by the front mirror is sent to a photodetector
$PD$, and the photodetection signal is demodulated before going to a spectrum analyzer.
The same signal is sent to a control system (not shown)
which provides a feedback signal applied to mirror $M_{1}$. The
noise of the system is studied by spectral analysis.
\label{Fig1}}
\end{figure}

In Fig.\ref{Fig1} it is represented the typical optical layout of the apparatus that
will be examined. Main features of the present model are: i) the multimode
description of the cavity field; ii) the inclusion of radiation pressure and
shot noise terms; iii) the description of suspensions and mirrors in terms
of mechanical modes.

Here and in the following $J=1,2$ labels the mirror, $x$ the axial and $q=y,z
$ the transverse coordinates. The analysis is focused on the fluctuations of
the $J$'s mirror orientation $(\delta \theta _{Jq})$ and displacements $%
(\delta x_{J},\delta \varepsilon _{Jq})$, combined in the parameters 
\begin{eqnarray}
\delta \psi _{J} &=&\left( -1\right) ^{J}2k^{\ell }\delta x_{J}  \nonumber \\
\delta \alpha _{Jq} &=&i\left( -1\right) ^{J}\sqrt{2}k^{\ell }w_{J}\left(
\delta \theta _{Jq}-\frac{\delta \varepsilon _{Jq}}{R_{J}}\right) 
\label{Mirror Coordinates}
\end{eqnarray}
with $k^{\ell }$ laser wavenumber, $w_{J}$ spot size, $R_{J}$ mirror
curvature radius and $\delta \theta _{Jq}=(\delta \vec{\Omega}_{J}\times 
\hat{x})\cdot \vec{q}$ depending on the rotation's angles $\delta \vec{\Omega%
}_{J}$. In addition the mirror vibrations are accounted for by
introducing matrices $\mathbf{\varsigma }_{Js}$ and functions $\delta
\varsigma _{Js}$ representing respectively profile and amplitude of the $s$%
-th vibrational mode of the $J$--th mirror.

The radiation pressure force and torques are linearized with respect to the
set $\left\{ \delta \alpha _{Jq}\right\} $ of transverse mirror coordinates
by introducing optical stiffness coefficients. Hence, the Fourier transforms 
$\left\{ \delta \widetilde{\alpha }_{Jq}\right\} $ satisfy Langevin
equations with driving forces proportional to these quantities. They become
important in proximity of the suspensions and mirror mechanical resonances.
These stiffness coefficients are in general frequency dependent for the presence
of the phase factor $e^{i\varpi \tau }$, with $\tau$ the cavity round trip.

A vector approach has been adopted by representing the amplitudes of the
excited cavity modes by a column vector $\mathbf{a}$ while the mismatch and
misalignment of the input beam is accounted for by a vector $\mathbf{v}$.
The various quantities $O$ used for describing the system dynamics (e.g. the
force acting on a given suspension mode) have been expressed in forms like $%
O=\mathbf{v}^{\dagger }\cdot \mathbf{O\cdot v}$ \ with $\mathbf{O}$ a matrix
representation of the quantity itself. In analogy with quantum mechanics 
$\mathbf{O}$ is seen as the matrix representation of an operator $\hat{O}$
corresponding to the quantity of interest, and $\mathbf{v}$ as the quantum
state of the resonator.

For evaluating spectral densities it has been introduced the symbol~$%
^{\ddagger }$ which is defined by its action on frequency dependent
quantities: 
\[
f^{\ddagger }\left( \varpi \right) \equiv f^{\ast }\left( -\varpi ^{\ast
}\right) \,,
\]
and the shorthands 
\begin{eqnarray*}
\Im \left\{ f\left( \varpi \right) \right\} =\frac{1}{2i}\left( f\left(
\varpi \right) -f^{\ddagger }\left( \varpi \right) \right) \\
\Re \left\{ f\left( \varpi \right) \right\} =\frac{1}{2}\left( f\left(
\varpi \right) +f^{\ddagger }\left( \varpi \right) \right) \,.
\end{eqnarray*}
The same $\ddagger $ applied to a frequency dependent matrix transforms $%
\mathbf{\ O}\left( \varpi \right) $ into $\mathbf{O}^{\ddagger }\left(
\varpi \right) =\mathbf{O}^{\dagger }\left( -\varpi ^{\ast }\right) $.

The summation symbol is omitted when applied to expressions containing a
repeated index.

The paper is organized in seven sections. Section~\ref{sec:Cavity-field} is
dedicated to the optical modes excited in a cavity with moving mirrors, and
to the susceptivities relative to the noise sources of the suspensions,
mirror vibrations and shot noise. The dynamic of the mechanical components
(mirrors and suspensions) is discussed in Sec.~\ref{sec:Radiation-pressure}
while the Drever-Pound and quadrant detector signals are analyzed in Sec.~%
\ref{sec:Error-signals}.

The results obtained in these sections are combined in Sec.~\ref
{sec:3D-model} where a five degrees of freedom model of the cavity,
including radiative pressure and torques, is presented. The model is
linearized for small misalignments and resonance enhanced effects are
discussed in Sec~\ref{sec:bipartite system} where the cavity is examined as
a bipartite system. In this context ponderomotive squeezing of the output
field and entanglement of two mirror modes are discussed. The manuscript is
completed by six mathematical appendices. The first three of them give the
expressions for the stiffness and the Drever-Pound signal matrices together
with their simplified expressions in case of small misalignment and
mismatch. The last ones are dedicated to thermal and shot-noise sources, and
their mutual correlations.

\section{Cavity field\label{sec:Cavity-field}}

A suspended cavity of length $L$ excited by a time harmonic field is
described by a superposition of Hermite-Gauss modes $u_{\lambda }\left( \vec{%
r},x\right) $ 
\[
e^{-i\omega ^{\ell }t}\sum_{\sigma =\pm }\exp \left[ i\sigma k^{\ell }\frac{%
r^{2}}{2R(x)}\right] a_{\lambda }^{\left( \sigma \right) }\left( x,t\right)
\,u_{\lambda }\left( \vec{r},x\right) \,.
\]
with $\omega ^{\ell }$ the laser frequency, $\sigma =+$ for a wave
travelling from mirror 1 toward 2 and $-$ contrariwise. The wavefront
curvature $R(x)$ is matched to the mirror's curvature: $R\left( 0\right)
=R_{1}<0$, $R\left( L\right) =R_{2}>0$. Each mode is labeled as usual by a
couple of integer numbers $(\lambda _{y},\lambda _{z})\equiv \lambda $. Here
and in the following x stands for the optical axis coordinate and $%
(y,z)\equiv \vec{r}$ \ are the two transverse ones. Each mode $u_{\lambda }$
is taken with a fixed normalization on the transverse section and without
phase factors, 
\[
u_{\lambda }(\vec{r})=u_{\lambda _{y}}(y)u_{\lambda _{z}}(z)\,,
\]
while the amplitudes are written as 
\[
a_{\lambda }^{\left( \sigma \right) }\left( x,t\right) =e^{i\sigma \left[
k^{\ell }x-\left( \lambda _{y}+\lambda _{z}+1\right) \phi \left( x\right) %
\right] }\left| a_{\lambda }^{\left( \sigma \right) }\left( x,t\right)
\right| 
\]
where $\phi \left( x\right) =\arctan \left( \frac{x-x_{0}}{b}\right) $ is
the phase delay of the Gaussian fundamental mode with respect to a plane
wave, $x_{0}$ being the distance of the waist from the input mirror and $b$
the confocal parameter. The field is propagated outside the resonator by
passing through the different optical components met on the way toward the
laser source and the photodetector which provides the error signal.

The laser beam incident on (input) mirror 1 has been split it in a classical
and in a quantum term: 
\begin{equation}
E^{in}\left( \vec{r},t\right) =e^{-i\omega ^{\ell }t}E\left( 1+\mu ^{\ell
}\left( t\right) \right) u^{in}\left( \vec{r}\right) +\delta \hat{a}^{SN}
\label{eq:FieldDecomposition}
\end{equation}
being 
\[
E=\sqrt{\frac{P}{\hbar \omega ^{\ell }}}=2.5\times 10^{9}\left( \frac{P}{1\,W%
}\right) ^{1/2}\left( \frac{\lambda ^{\ell }}{1\,\mu }\right) ^{1/2}\text{Hz}%
^{1/2} 
\]
the mean amplitude, and $\mu ^{\ell }$ the relative amplitude fluctuations.
Effects of the laser linewidths have been ignored.

Misalignment and mismatch effects between the input beam and the cavity are
taken into account by writing $u^{in}\left( \vec{r}\right) $ as a
superposition of cavity modes, namely 
\begin{equation}
u^{in}\left( \vec{r}\right) =v_{\lambda }\,u_{\lambda }\left( \vec{r}\right)
\,.  \label{mode expansion}
\end{equation}
The structure of the expansion coefficients is factorized in a product of
Hermite polynomials 
\begin{equation}
v_{\lambda }\propto \frac{\delta _{y}\delta _{z}H_{\lambda _{y}}\left( \text{%
v}_{y}\frac{\delta _{y}^{2}-1}{\delta _{y}\sqrt{2}}\right) H_{\lambda
_{z}}\left( \text{v}_{z}\frac{\delta _{z}^{2}-1}{\delta _{z}\sqrt{2}}\right) 
}{\sqrt{2^{\lambda _{y}+\lambda _{z}}\lambda _{y}!\lambda _{z}!}}\,
\label{misalignment}
\end{equation}
depending on the misalignment v$_{q}$ and mismatch $\delta _{q}$ parameters
defined respectively by 
\begin{eqnarray*}
\text{v}_{q} &=&-i\frac{k^{\ell }w_{1}}{\sqrt{2}}\left( \theta _{q}-\frac{%
\varepsilon _{q}}{Q_{q}}\right) \\
\delta _{q} &=&\sqrt{1+i\frac{2}{k^{\ell }w_{1}^{2}}\frac{Q_{q}Q_{1}^{\ast }%
}{\left( Q_{q}-Q_{1}^{\ast }\right) }}\,.
\end{eqnarray*}
For a perfect matching $\text{v}=0$ and $\delta =0$. Here $Q_{1}$ is the
complex curvature radius of the cavity mode evaluated at the input mirror,
while $Q_{q}$, $\theta _{q}$ and $\varepsilon _{q}$ stand for the curvature
radius, angular and transverse misalignment of the input beam.

The modal expansion~(\ref{mode expansion}) will be used in the following for
representing the cavity fields in correspondence of the two mirrors as
column vectors $\mathbf{v}$ with components $v_{\lambda }.$ So doing the
multiplication of $u\left( \vec{r}\right) $ by a function $w\left( \vec{r}%
\right) $ will be represented by the product $\ \mathbf{w\cdot v}$ of $%
\mathbf{v}$ by a matrix $\mathbf{w}$.

The coupling of the cavity with the universe modes through the partially
transmitting mirrors introduces the quantum noise contribution $\delta 
\hat{a}^{SN}\left( \mathbf{r,}t\right) $ of Eq.~(\ref{eq:FieldDecomposition}%
) 
\[
\left[ \delta \hat{a}^{SN}\left( \mathbf{r,}t\right) ,\delta \hat{a}%
^{SN\dagger }\left( \mathbf{r}^{\prime }\mathbf{,}t^{\prime }\right) \right]
=\delta ^{\left( 3\right) }\left( \mathbf{r}-\mathbf{r}^{\prime }\right)
\delta \left( t-t^{\prime }\right) 
\]
It can be expanded as a superposition of delta-correlated operators $\delta 
\hat{a}_{\lambda }^{SN}\left( t\right) $, 
\begin{equation}
\delta \hat{a}^{SN}\left( \mathbf{r,}t\right) =\delta \hat{a}_{\lambda
}^{SN}\left( t\right) u_{\lambda }\left( \mathbf{r}\right) \,.
\label{shot-noise contributions}
\end{equation}

Before arriving at the mirror the excitation beam is passed through a phase
modulator represented by the phase factor 
\begin{equation}
F=e^{iM\sin \left( \Lambda t\right) }=e^{-ip\Lambda t}J_{p}\left( M\right)
\label{Ffunction}
\end{equation}
with $J_{p}\left( M\right) $ the p-th Bessel function of argument $M$. The
input modulation $F$ modifies the laser excited amplitude $a_{\lambda
}^{\ell }$ into a sum of harmonics varying on the time scale of the
suspension fluctuations, 
\begin{equation}
a_{\lambda }=J_{p}e^{-ip\Lambda t}a_{\lambda p}  \label{ap}
\end{equation}
while leaving the noise unaffected.

\subsection{Cavity fluctuations}

Owing to the fluctuations of the suspensions the mirror orientations change
slowly in time by undergoing torsions $\delta \Omega _{Jz}\left( t\right)$,
tiltings $\delta \Omega _{Jy}\left( t\right) $ and transverse displacements $%
\delta \vec{\varepsilon}_{J}\left( t\right) $. The mirror can rotate also
around the optical axis, but this motion is uncoupled to the cavity field in
the linear approximation.

The mirror motions separate into fluctuating and average components, the
latter ones setting the reference frame for the vector representation. So
doing the average misalignment and displacements will be included in those
relative to the input beam, which will be represented by a unit amplitude
vector $\mathbf{v}_J$ 
\[
\mathbf{v}_{1}=\mathbf{v\;,\;v}_{2}=\mathbf{\Phi }^{\frac{1}{2}}\cdot 
\mathbf{v} 
\]
with $\mathbf{\Phi }$ a diagonal matrix of components $\Phi _{\lambda
}=e^{-i2\left( \lambda _{y}+\lambda _{z}+1\right) \phi _{G}}$ and $\phi
_{G}=\phi \left( L\right) -\phi \left( 0\right) $ the single-trip phase
delay of the Gaussian fundamental mode. Accordingly in the following the
parameters $\delta \alpha _{Jq}$ (see Eq. (\ref{Mirror Coordinates})
will be small fluctuating quantities.

The reflection at mirror $1$ induces the transformation $u_{\lambda
}a_{\lambda }^{\left( -\right) }=\mathfrak{r}_{1}u_{\lambda }a_{\lambda
}^{\left( +\right) }$ with $\mathfrak{r}_{1}\left( t\right) $ the phase
factor 
\begin{widetext}
\begin{equation}
\mathfrak{r}_{1} =r_{1}\exp \left[ -ik^{\ell }\frac{\delta \varepsilon
_{1}^{2}}{R_{1}}-i2k^{\ell }\delta x_{1}-i2k^{\ell }\delta
u_{1}^{DEF}+i2k^{\ell }\left( \delta \vec{\Omega}_{1}\times \hat{x}-\frac{%
\delta \vec{\varepsilon}_{1}}{R_{1}}\right) \cdot \vec{r}\right] \,.
\label{reflection}
\end{equation}
\end{widetext}
Here $\delta x_{1}\left( t\right) $ is the deviation of the center from the
positions at rest $\left( x_{1}\left( t\right) =0+\delta x_{1}\left(
t\right) \right) $. $\delta u_{1}^{DEF}\left( \vec{r},t\right) $ is the tiny
deformation of the mirror surface represented by the matrix $\delta \mathbf{%
\ \varsigma }_{1}\left( t\right) $ of components 
\begin{equation}
\delta \varsigma _{1\lambda \lambda ^{\prime }}\left( t\right) =2k^{\ell
}\int u_{\lambda }\left( \vec{r}\right) \delta u_{1}^{DEF}\left( \vec{r}%
,t\right) u_{\lambda ^{\prime }}\left( \vec{r}\right) d^{2}\vec{r}\;
\label{scattering matrix}
\end{equation}
Expanding further $\delta u_{1}^{DEF}$ into mirror modes \cite
{Saulson,Hadjar} $\delta \mathbf{\varsigma }_{1}\left( t\right) $ becomes a
superposition 
\begin{equation}
\delta \mathbf{\varsigma }_{1}\left( t\right) =\delta \varsigma _{1s}\left(
t\right) \mathbf{\varsigma }_{1s}  \label{deformation modes}
\end{equation}
of matrices $\mathbf{\varsigma }_{1s}$ times fluctuating amplitudes $\delta
\varsigma _{1s}\left( t\right) $ driven by radiation pressure and thermal
noise.

Although the frequencies of the mirror acoustic modes are very large, the
tails of their spectra contribute to the low frequency thermal noise of the
interferometers as recently reported by~\cite{Numata}. Levin~ \cite{Levin}
has approximated, at very low frequency, the many mode profiles with the
steady-state mirror surface deformation $\delta u_{1}^{DEF}\left( \vec{r}%
\right) $ (in matrix form $\mathbf{\varsigma }_{1}^{L}$) under the action of
the incident beam (positive for a compression), by replacing Eq.(\ref
{deformation modes}) with 
\begin{equation}
\delta \mathbf{\varsigma }_{1}\left( t\right) =\mathbf{\varsigma }%
_{1}^{L}\delta \varsigma _{1}^{L}\left( t\right) 
\label{low frequency deformation}
\end{equation}
$\delta \varsigma _{1}^{L}\left( t\right) $ being a stochastic process, (see
Eq.(\ref{spectrum mirror noise})).

Accordingly, ignoring the quadratic expression $k^{\ell }\delta \varepsilon
_{1}^{2}/R_{1}$ the phase factor $\mathfrak{r}_{1}$ (Eq.(\ref{reflection}))
is represented in vector form by 
\begin{equation}
r_{1}e^{-i\left( 2k^{\ell }\delta x_{1}+\delta \mathbf{\varsigma }%
_{1}\right) }\cdot \mathbf{D}_{1}\left( -\delta \mathbf{\alpha }_{1}\right)
\label{displacement}
\end{equation}
with $\delta \mathbf{\alpha }_{1}=\left( \delta \alpha _{1y},\delta \alpha
_{1z}\right) $ the combination of rotation and displacement defined by Eq.(%
\ref{Mirror Coordinates}) and $\mathbf{D}_{1}$ the displacement operator 
\[
\mathbf{D}_{1}\left( -\delta \alpha _{1}\right) =\exp \left( -\delta \alpha
_{1q}\mathbf{B}_{q}^{\dagger }+\delta \alpha _{1q}^{\ast }\mathbf{B}%
_{q}\right) 
\]
acting on the functions of the transverse coordinates. The operators $%
\mathbf{B}_{y}$ and $\mathbf{B}_{z}$ act on the mode functions $u_{\lambda }$
as typical annihilation operators, $\mathbf{B}u_{n}=\sqrt{n}u_{n-1}$, and
this is the reason why $\mathbf{D}$ has been called a displacement operator.

Next, the propagation from the input mirror to the opposite one is described
by 
\begin{equation}
e^{ik^{\ell }L}\left( \hat{D}_{t}\mathbf{\Phi }\right) ^{\frac{1}{2}}
\label{propagation}
\end{equation}
with $\hat{D}_{t}=e^{-\tau \frac{d}{dt}}$ the delay operator by the cavity
round-trip time $\tau $. Next combining (\ref{displacement}) with (\ref
{propagation}) a round trip is represented by 
\begin{widetext}
\[
R\,e^{-i\psi -i\delta \psi _{1,cav}}\left( \hat{D}_{t}\mathbf{\Phi }\right)
^{\frac{1}{2}}\cdot e^{-i\delta \mathbf{\varsigma }_{2}}\cdot \mathbf{D}%
_{2}\left( -\delta \mathbf{\alpha }_{2}\right) \left( \hat{D}_{t}\mathbf{%
\Phi }\right) ^{\frac{1}{2}}\cdot e^{-i\delta \mathbf{\varsigma }_{1}}\cdot 
\mathbf{D}_{1}\left( -\delta \mathbf{\alpha }_{1}\right)
\]
\end{widetext}
where $\psi $ is the detuning phase ($\psi >0$ for a cavity shorter than the
closest resonance length), $R=r_{1}r_{2}=e^{-\mathcal{F}/\pi }$ with $%
\mathcal{F}$ the cavity finesse, and $\delta \psi _{1,cav}$ the accumulated
phase shift, positive for decreasing cavity length, 
\[
\delta \psi _{1,cav}\left( t\right) =\delta \psi _{1}\left( t-\tau \right)
+\delta \psi _{2}\left( t-\frac{\tau }{2}\right) 
\]
with $\delta \psi _{J}=-\left( -1\right) ^{J}2k^{\ell }\delta x_{J}.$ Next,
in view of the smallness of $\delta \psi _{1,cav},\delta \mathbf{\alpha }_{J}
$ and $\delta \mathbf{\varsigma }_{J}$ $\mathbf{D}_{1,2}$ and $e^{-i\delta
\psi _{1,cav}}$ can be linearized thus obtaining for the round-trip
transformation 
\begin{equation}
e^{-i\psi }R\mathbf{\Phi }\left( \hat{D}_{t}-i\mathfrak{X}\cdot %
\mathfrak{\delta \alpha }_{J,cav}-i\delta \mathbf{\varsigma }_{J,cav}\right)
\,.  \label{eq:roundtrip compact}
\end{equation}
Here $\mathfrak{X}\cdot \mathfrak{\delta \alpha }_{J,cav}$ indicates the sum 
$\mathfrak{X}_{i}\left( \mathfrak{\delta
\alpha }_{J,cav}\right) _{\mathfrak{i}}$ and two vectors 
\begin{eqnarray}
\mathfrak{X} &=&\left( 1\mathbf{,X}_{y},\mathbf{X}_{z},\mathbf{Y}_{y},%
\mathbf{Y}_{z}\right) \label{eq:XSUS}\\
\mathfrak{\delta \alpha }_{J,cav} &=&\left( \delta \psi _{J,cav},\delta
\alpha _{Jy,cav}^{\prime \prime },\delta \alpha _{Jz,cav}^{\prime \prime
},\delta \alpha _{Jy,cav}^{\prime },\delta \alpha _{Jz,cav}^{\prime }\right)
\nonumber
\end{eqnarray}
collect the phase quadratures $\mathbf{X}_{q}=\mathbf{B}_{q}+\mathbf{B}%
_{q}^{\dagger }$, $\mathbf{Y}_{q}=i\left( \mathbf{B}_{q}-\mathbf{B}%
_{q}^{\dagger }\right) $ and the combinations 
\[
\delta \alpha _{1q,cav}\left( t\right) =\delta \alpha _{1q}\left( t-\tau
\right) +e^{i\phi _{G}}\delta \alpha _{2q}\left( t-\frac{\tau }{2}\right) 
\]
($\alpha ^{\prime }$ and $\alpha ^{\prime \prime }$ are the real and
imaginary part of $\alpha $ respectively).

Analogously for $\delta \mathbf{\varsigma }_{J,cav}$ 
\[
\delta \mathbf{\varsigma }_{1,cav}\left( t\right) =\mathbf{\varsigma }%
_{Js}\delta \varsigma _{Js}\left( t-\tau \right) +\mathbf{\Phi }^{-\frac{1}{2%
}}\cdot \mathbf{\varsigma }_{2s^{\prime }}\cdot \mathbf{\Phi }^{\frac{1}{2}%
}\,\delta \varsigma _{2s^{\prime }}\left( t-\frac{\tau }{2}\right) 
\]

The amplitude $a_{\lambda _{y}\lambda _{z}}^{\left( \sigma \right) }\left(
t\right) $ of the $\lambda _{y}\lambda _{z}$-th mode is propagated back and
forth the cavity. The fraction $t_{1}$ is injected into the Fabry Perot
through mirror $1$ at time $t$, propagates toward and is reflected by mirror 
$2$ at $t+\frac{\tau }{2}$ and again by $1$ at $t$. Hence, summing over the
sequence of rund-trips, the field $\mathbf{a}_{J}$ incident on the $J$-th
mirror reads 
\begin{equation}
\mathbf{a}_{J}=\mathcal{E}\left( 1+\mu ^{\ell }\right) \mathbf{\hat{G}}_{J}%
\mathbf{\cdot v}_{J}F+\delta \mathbf{\hat{a}}^{SN}
\label{total laser contribution}
\end{equation}
with $\mathcal{E}=t_{1}E$ and 
\[
\mathbf{\hat{G}}_{J}=\frac{1}{1-Re^{-i\psi }\mathbf{\Phi \cdot }\left( \hat{D%
}_{t}-i\mathfrak{X}\cdot \mathfrak{\delta \alpha }_{J,cav}-i\delta \mathbf{%
\varsigma }_{J,cav}\right) }
\]

For very small $\mathfrak{\delta \alpha }_{J,cav}$ and $\delta \mathbf{\
\varsigma }_{J,cav}$ first-order perturbation theory can be used. On the
other hand assuming for $u_{\lambda }$ either Hermite or Laguerre-Gauss
modes the various terms of the perturbation $\mathfrak{X}\cdot %
\mathfrak{\delta \alpha }_{J,cav}+\delta \mathbf{\varsigma }_{J,cav}$ do not
couple the respective degenerate modes. Hence, the Green operator $\mathbf{%
\hat{G}}_{J}$ can be expressed as 
\begin{equation}
\mathbf{\hat{G}}_{J}\simeq \mathbf{\hat{G}-}i\tilde{\mathfrak{G}}\cdot %
\mathfrak{\delta \alpha }_{J,cav}\mathbf{-}i\delta \mathbf{\hat{G}}^{DEF}
\label{G small}
\end{equation}
where the first term on the right is a static propagator, the second the
contribution of the linearized motion of the mirrors and the third one
describes the mirror deformations,
\begin{eqnarray}
\mathbf{\hat{G}} &=&\left( 1-Re^{-i\psi }\hat{D}_{t}\mathbf{\Phi }\right)
^{-1}  \nonumber \\
\mathfrak{G} &=&e^{-i\psi }R\mathbf{\hat{G}}\cdot \mathbf{\Phi \cdot }%
\mathfrak{X}\cdot \mathbf{\hat{G}}  \nonumber \\
\delta \mathbf{\hat{G}}^{DEF} &=&e^{-i\psi }R\mathbf{\hat{G}}\cdot \mathbf{%
\Phi \cdot }\delta \mathbf{\varsigma }_{J,cav}\cdot \mathbf{\hat{G}}
\label{approximate Green}
\end{eqnarray}

Next, the contributions of the shot noises entering the cavity through
mirror $J$ has been split as $\delta \mathbf{\hat{a}}^{SN}=\delta \mathbf{%
\hat{a}}_{1}^{SN}+t_{2}t_{1}^{-1}\delta \mathbf{\hat{a}}_{2}^{SN}$, so that
the same approximation of Eq.~(\ref{approximate Green}) applies and 
\begin{equation}
\mathbf{a}_{J}=\mathbf{a}_{0,J}+\delta \mathbf{a}_{J}+\delta \mathbf{\hat{a}}%
^{SN}  \label{approx-laser-contrib}
\end{equation}
Here $\delta \mathbf{a}_{J}$ is fluctuating with the cavity geometry and
laser intensity, while $\mathbf{a}_{0,J}$ does not depend on it and on shot
noise,
\begin{eqnarray}
\mathbf{a}_{0,J} &\approx &\mathcal{E}\mathbf{\hat{G}\cdot v}_{J}F  \nonumber
\\
\delta \mathbf{a}_{J} &\approx &\mathcal{E}\left( \mu ^{\ell }\mathbf{\hat{G}%
-}i\mathfrak{G}\cdot \mathfrak{\delta \alpha }_{J,cav}\mathbf{-}i\delta 
\mathbf{\hat{G}}^{DEF}\right) \mathbf{\cdot v}_{J}F\,  \nonumber \\
\delta \mathbf{\hat{a}}^{SN} &=&t_{1}\mathbf{\hat{G}\cdot }\left( \delta 
\mathbf{\hat{a}}_{1}^{SN}+\frac{t_{2}}{t_{1}}\delta \mathbf{\hat{a}}%
_{2}^{SN}\right) .  \label{cavity field}
\end{eqnarray}

Further, the relation $\hat{D}_{t}e^{-ip\Lambda t}=e^{-ip\Lambda
t}e^{ip\Lambda \tau }\hat{D}_{t}$ implies $\mathbf{\hat{G}}e^{-ip\Lambda
t}=e^{-ip\Lambda t}\mathbf{\hat{G}}_{p}$ with the suffix $p$ indicating that 
$R$ has been replaced by $R_{p}=e^{ip\Lambda \tau }R.$ Then, the factor $%
e^{-ip\Lambda t}$ contained in the function $F$ (see Eq. \ref{Ffunction}) 
can be displaced from the
right to the left side of the above expressions by adding the suffix $p$ to
the various Green's functions. Hence 
\[
\mathbf{a}_{J}=e^{-ip\Lambda t}\left( \mathbf{a}_{0,Jp}+\delta \mathbf{%
\hat{a}}_{Jp}\right) +\delta \mathbf{\hat{a}}^{SN} 
\]
where 
\begin{eqnarray}
\mathbf{a}_{0,Jp} &=&\mathcal{E}J_{p}\mathbf{G}_{p}\mathbf{\cdot v}_{J} \\
\delta \mathbf{a}_{Jp} &=&e^{-ip\Lambda t}\mathcal{E}J_{p}\left( \mu ^{\ell }%
\mathbf{G}_{p}-i\mathfrak{G}_{p}\cdot \mathfrak{\delta \alpha }_{J,cav}%
\mathbf{-}i\delta \mathbf{\hat{G}}_{p}^{DEF}\right) \mathbf{\cdot v}_{J} 
\nonumber
\end{eqnarray}
Analogously for the output field~\cite{Collet} 
\begin{eqnarray}
\mathbf{a}_{0,Jp}^{OUT} &=&t_{1}\mathcal{E}J_{p}\mathbf{G}_{p}^{OUT}\mathbf{%
\cdot v}_{1}  \nonumber \\
\delta \mathbf{a}_{Jp}^{OUT} &=&t_{1}\mathcal{E}J_{p}\left( \mathbf{G}%
_{p}^{OUT}\mu ^{\ell }-i\mathfrak{G}_{p}\cdot \mathfrak{\delta \alpha }%
_{1,cav}\right.  \nonumber \\
&&\hspace{1in}\left. -i\delta \mathbf{\hat{G}}_{p}^{DEF}\right) \mathbf{%
\cdot v}_{1}  \nonumber \\
\delta \mathbf{\hat{a}}^{OUT\;SN} &=&t_{1}^{2}\left( \mathbf{\hat{G}}^{OUT}%
\mathbf{\cdot }\delta \mathbf{\hat{a}}_{1}^{SN}+\frac{t_{2}}{t_{1}}\mathbf{%
\hat{G}\cdot }\delta \mathbf{\hat{a}}_{2}^{SN}\right)  \label{aout}
\end{eqnarray}
where $\mathbf{\hat{G}}^{OUT}=\mathbf{\hat{G}}-r_{1}/t_{1}^{2}$.

\section{Radiation pressure and torque \label{sec:Radiation-pressure}}

Bouncing back and forth the two mirrors the laser and shot noise fields
exert a radiation pressure resulting in an axial force directed along the
optic axis $\hat{x}$ and a torque parallel to their surfaces, proportional
to the total intensity $\mathbf{a}_{J}^{\dagger }\mathbf{\cdot a}_{J}$ and
moments $\mathbf{a}_{J}^{\dagger }\cdot \mathbf{X}_{q}\cdot \mathbf{a}_{J}$.
They split into classical $\mathcal{F}_{J\;rp}$, $\mathcal{T}_{J\;rp}$ and
quantum $\mathcal{F}_{rp}^{SN}$, $\mathcal{T}_{J\;rp}^{SN}$ components
respectively given by 
\begin{eqnarray}
\mathcal{F}_{J\;rp} &=&\left( -1\right) ^{J}\mathcal{E}^{2}\,2\frak{R}%
_{J}\,\hbar k^{\ell }\left( F_{0,J}+\delta F_{J}\right) \hat{x}
\label{eq:FT1} \\
\mathcal{T}_{J\;rp} &=&\left( -1\right) ^{J}\mathcal{E}^{2}\,2\frak{R}%
_{J}\,\hbar k^{\ell }\frac{w_{J}}{\sqrt{2}}\left( T_{0,Jq}+\delta
T_{Jq}\right) \hat{q}\times \hat{x}  \nonumber
\end{eqnarray}
and 
\begin{eqnarray}
\mathcal{F}_{J\;rp}^{SN} &\equiv &\left( -1\right) ^{J}\mathcal{E\,}2\frak{R}%
_{J}\,\hbar k^{\ell }\hat{X}_{J\psi }^{SN}\hat{x}  \nonumber \\
\mathcal{T}_{J\;rp}^{SN} &\equiv &\left( -1\right) ^{J}\mathcal{E\,}2\frak{R}%
_{J}\,\hbar k^{\ell }\frac{w_{J}}{\sqrt{2}}\hat{X}_{J\theta q}^{SN}\hat{q}%
\times \hat{x}  \label{eq:FT2}
\end{eqnarray}
where $\frak{R}_{J}=\left| r_{J} \right|^{2}+\frac{1}{2}A_{J}$
with $A_{J}$ the 
$J$--th mirror power absorption.

$F_{0,J}$ and $\delta F_{J}$ indicate the contributions of $\mathbf{a}%
_{0,J}^{\dagger }\mathbf{\cdot a}_{0,J}$ and $\mathbf{a}_{0,J}^{\dagger
}\cdot \delta \mathbf{a}_{J}+H.c.$ and analogously for $T_{0,Jq},\delta
T_{Jq}$. $F_{0,J},T_{0,Jq}$ split in turn into time constant terms $\bar{F}%
_{0}$, $\bar{T}_{0,Jq},$ balanced by the stabilization system of the
apparatus, and small terms $\delta F_{0,J},\delta T_{0,Jq}$ oscillating at
multiples of $\Lambda $. Being $\Lambda $ typically of the order of some MHz
these contributions can be ignored.

For a stabilized resonator $\hat{\mathbf{G}}_{J}$ is represented as in~(\ref
{G small}) so that $\delta F_{J}$ and $\delta T_{Jq}$ reduce in the
frequency domain respectively to 
\begin{eqnarray}
\delta \tilde{F}_{J} &=&\bar{F}_{0,J}\tilde{\mu}^{\ell }+\tilde{\mathfrak{F}}%
_{J}\cdot \mathfrak{\delta \tilde{\alpha}}_{J,cav}+\delta \tilde{F}%
_{J,cav}^{DEF}  \nonumber \\
\delta \tilde{T}_{Jq} &=&\bar{T}_{0,Jq}\tilde{\mu}^{\ell }+\tilde{%
\mathfrak{T}}_{Jq}\cdot \mathfrak{\delta}\tilde{\mathfrak{\alpha}}%
_{J,cav}+\delta \tilde{T}_{Jq,cav}^{DEF}  \label{eq:RPtorque}
\end{eqnarray}
The three pieces of Eqs.~(\ref{eq:RPtorque}) represent, in the given order,
the contributions of the fluctuation of laser intensity, mirror
displacements, rotations and surface deformations to the radiation pressure
forces and torques.

Being the suspension characteristic frequencies generally smaller than the
mirror modes resonances \cite{Hadjar}, the deformations are described by a
single matrix (see Eq. (\ref{low frequency deformation})).

The vectors $\tilde{\mathfrak{F}}_{J}=\left( F_{J\psi
},F_{JXq},F_{JYq}\right) $ and $\tilde{\mathfrak{T}}_{Jq}=\left( T_{J\psi
q},T_{JqXq^{\prime }},T_{JqYq^{\prime }}\right) $ contain five
proportionality constants between the forces (the torques) and the
coordinates $\left( \mathfrak{\delta}\tilde{\mathfrak{\alpha}}%
_{J,cav}\right) $ which parametrize the mirror's displacement, so they are
stiffness coefficients. $\bar{F}_{0,J}$, $\tilde{\mathfrak{F}}_{J}$, $\delta 
\tilde{F}_{J,cav}^{DEF}$ and $\bar{T}_{0,Jq}$, $\tilde{\mathfrak{T}}_{Jq}$, $%
\delta \tilde{T}_{Jq,cav}^{DEF}$ depend on the steady-state amplitudes of
the cavity modes, represented by the vector $\mathbf{v}_{J}$, 
\begin{eqnarray}
\bar{O}_{0,J} &=&\mathbf{v}_{J}^{\dagger }\cdot \mathbf{\bar{O}}_{0}\cdot 
\mathbf{v}_{J}  \nonumber \\
\tilde{O}_{J} &=&\mathbf{v}_{J}^{\dagger }\cdot \mathbf{\tilde{O}}\cdot 
\mathbf{v}_{J}  \label{matrix representations}
\end{eqnarray}
with $\bar{O}_{0,J}=\bar{F}_{0,J}$, $\bar{T}_{0,Jq}$ and $\tilde{O}=\tilde{%
\mathfrak{F}}_{J}$, $\delta \tilde{F}_{J,cav}^{DEF}$, $\mathfrak{T}_{Jq}$
and $\delta \tilde{T}_{Jq,cav}^{DEF}$ Matrices $\mathbf{\bar{F}}_{0},\delta 
\tilde{\mathbf{F}}_{J,cav}^{DEF}$ and $\mathbf{\bar{T}}_{0,q},\delta \tilde{%
\mathbf{T}}_{Jq,cav}^{DEF}$ are reported in Appendix A (Eqs. (\ref{F0T0},\ref
{FTDEF}) ) while $\tilde{\mathfrak{F}},\tilde{\mathfrak{T}}_{q}$ are
collections of five matrices (Eq.~(\ref{stiffness matrices})). They depend
on Green's matrices (Eqs. (\ref{stiffness terms},\ref{FTDEFM})), and through
them on frequency and detuning, closeness of cavity modes with respect to
linewidth and phase modulation depth. The frequency dependence is due to the
factor $e^{i\varpi \tau }$ appearing in different fashions in $\tilde{%
\mathbf{G}}_{p}$, $\tilde{\mathbf{G}}_{p}^{OUT}$, $\tilde{\mathbf{G}}_{p}$.

Eventually, the shot-noise contributions (Eq.~(\ref{eq:FT2})) are expressed
by 
\[
\mathbf{\hat{X}}_{Ji}^{SN}=t_{1}^{-1}J_{p}\mathbf{v}_{J}^{\dagger }\cdot 
\mathbf{G}_{p}^{\dagger }\mathbf{\cdot X}_{i}\cdot \delta \mathbf{\hat{a}}%
^{SN}e^{ip\Lambda t}+H.c.
\]
with $i\in (\psi ,\theta _{y},\theta _{z})$ and take in the frequency domain
the form 
\begin{equation}
\tilde{\mathbf{X}}_{Ji}^{SN}=t_{1}^{-1}2J_{p}\Re \left\{ \mathbf{v}%
_{J}^{\dagger }\cdot \mathbf{G}_{p}^{\dagger }\mathbf{\cdot X}_{i}\cdot
\delta \mathbf{\tilde{a}}_{p}^{SN}\right\}   \label{suspension shot noise}
\end{equation}
Finally, on the $J$'s mirror mode act the forces, 
\begin{eqnarray}
\mathcal{F}_{Js\;rp}^{DEF} &=&\left( -1\right) ^{J}\mathcal{E}^{2}\,2\frak{R}%
_{J}\,\hbar k^{\ell }  \nonumber \\
&&\hspace{0.5in}\left( F_{0,Js}^{DEF}+\delta F_{Js,cav}^{DEF}\right) \hat{x}
\nonumber \\
\mathcal{F}_{Js\;rp}^{DEF\;SN} &\equiv &\left( -1\right) ^{J}\mathcal{E\,}2%
\frak{R}_{J}\hbar k^{\ell }\hat{X}_{Js}^{DEF\;SN}\hat{x}
\label{mirror modes force}
\end{eqnarray}
where 
\[
\delta \tilde{F}_{Js,cav}^{DEF}=e^{i\varpi \tau }\tilde{F}_{JsJs^{\prime
}}^{DEF}\delta \tilde{\varsigma}_{Js^{\prime }}+e^{i\varpi \tau /2}\tilde{F}%
_{Js\bar{J}s^{\prime }}^{DEF}\delta \tilde{\varsigma}_{\bar{J}s^{\prime }}.
\]
Here 
\begin{equation}
\tilde{F}_{JsJ^{\prime }s^{\prime }}^{DEF}=\mathbf{v}_{J}^{\dagger }\cdot 
\tilde{\mathbf{F}}_{JsJ^{\prime }s^{\prime }}^{DEF}\cdot \mathbf{v}_{J}
\label{stiffness deformation}
\end{equation}
is the force acting on the $Js$-mode due to the deformations of the mirror
surfaces. In this case the force does not factorize as for the suspension
modes. $\tilde{\mathbf{F}}_{JsJ^{\prime }s^{\prime }}^{DEF}$ (Eq. (\ref
{stiffnes mirror modes})) represent the effects of the vibrations of the
modes $J^{\prime }s^{\prime }$ on the $Js$ one.

Next, the shot-noise force is given by 
\begin{equation}
\tilde{X}_{Js}^{DEF\;SN}=t_{1}^{-1}2J_{p}\Re \left\{ \mathbf{v}_{J}^{\dagger
}\cdot \mathbf{G}_{p}^{\dagger }\mathbf{\cdot \varsigma }_{Js}\cdot \delta 
\mathbf{\tilde{a}}_{p}^{SN}\right\}  \label{mirror mirror forces}
\end{equation}

\begin{figure}
\includegraphics[width=7cm]{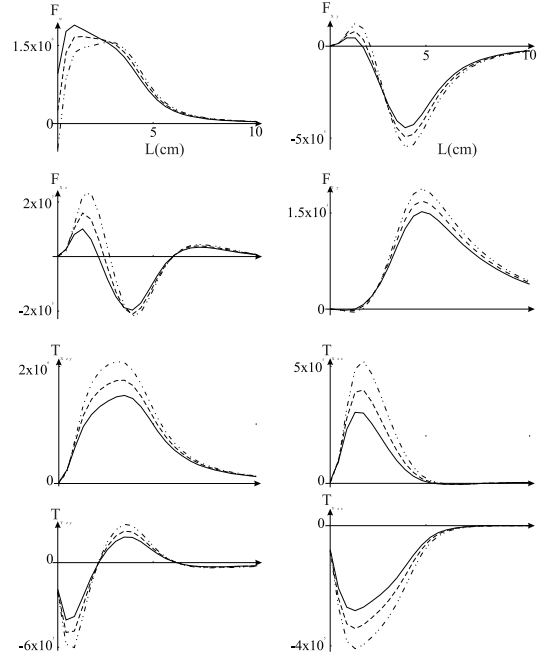}
\caption{
Axial $F_{\psi }$, $F_{1Xy}$, $F_{1Xz}$ $F_{1Yy}$ and angular $%
T_{1zXy}$, $T_{1zXz}$, $T_{1zYy}$, $T_{1zYz}$ stiffness coefficients vs.
length L of a symmetric cavity for angular misalignments $\theta _{y}=.01$,$%
\theta _{z}=.1$ mrad and detunings $\psi =.1,.2,.3$ $\pi /\mathcal{F}$ . The
round-trip phase factor $e^{i\varpi \tau }$ has been ignored.
\label{Fig2}}
\end{figure}

In Fig. \ref{Fig2} the optically induced stiffness coefficients have been plotted for
a set of detunings and angular misalignments, in an almost concentric cavity
having a finesse $\mathcal{F}=500$ and output spot sizes of $2\times
10^{-3}\,m$. Being close to the concentric configuration also the stiffness
coefficients $F_{X/Yq},T_{zXy},T_{yYz}$ become comparable with $F_{\psi
},T_{JqX/Yq}$ for cavity axis misaligned by $\theta _{y}=10^{-2}\;rad,$ $%
\theta _{z}=10^{-1}\;rad.$ The signs of the stiffness coefficients may have
important consequences on the mechanical stability, as discussed by several
authors for plane-parallel and concave mirrors \cite{Deruelle,Sidles}.

\subsection{\textbf{Small misalignment and mismatch}}

In the limit of small misalignment and mismatch the vector $\mathbf{v}_{1}$
(Eq. (\ref{misalignment})) reduces to $\simeq \mathbf{1+}\delta \mathbf{v=}%
\left\{ 1,\left( \delta \text{v}_{y}\;,\delta \text{v}_{z}\right)
,0,...,0\right\} $ with 
\[
\delta \text{v}_{q}=-\frac{\sqrt{2}}{w_{1}}\frac{Q^{\ast }Q_{q}}{Q^{\ast
}-Q_{q}}\left( \theta _{q}-\frac{\varepsilon _{q}}{R_{1}}\right) 
\]
For mirror 2 $\mathbf{v}_{q}$ \ is multiplied by $e^{-i\phi _{G}}$.
Splitting forces and torques in 0-th and 1-st order terms in these
misalignment parameters $\delta \tilde{F}_{J}$ and $\delta \tilde{T}_{Jq}$
of Eqs. (\ref{eq:RPtorque}) take the simpler forms, 
\begin{eqnarray}
\delta \tilde{F}_{J} &=&\delta \tilde{F}_{J}^{\left( 0\right) }+\delta 
\tilde{F}_{J}^{\left( 1\right) }  \nonumber \\
\delta \tilde{T}_{Jq} &=&\delta \tilde{T}_{Jq}^{\left( 0\right) }+\delta 
\tilde{T}_{Jq}^{\left( 1\right) }  \label{rad press bis}
\end{eqnarray}
where 
\begin{eqnarray*}
\delta \tilde{F}_{J}^{\left( 0\right) } &=&\bar{F}_{0,J}\tilde{\mu}^{\ell }+%
\tilde{F}_{\psi }\delta \tilde{\psi}_{J,cav}+\delta \tilde{F}_{J,cav}^{DEF}
\\
\delta \tilde{F}_{J}^{\left( 1\right) } &=&\delta \tilde{F}_{JXq}\delta 
\tilde{\alpha}_{Jq,cav\,}^{\prime \prime }+\delta \tilde{F}_{JYq}\delta 
\tilde{\alpha}_{Jq,cav\,}^{\prime } \\
\delta \tilde{T}_{Jq}^{\left( 0\right) } &=&\tilde{T}_{X}\delta \tilde{\alpha%
}_{Jq,cav\,}^{\prime \prime }+\tilde{T}_{Y}\delta \tilde{\alpha}%
_{Jq,cav\,}^{\prime } \\
\delta \tilde{T}_{Jq}^{\left( 1\right) } &=&\delta \tilde{T}_{Jq}\delta 
\tilde{\psi}_{J,cav}
\end{eqnarray*}
with
\begin{eqnarray*}
\delta \tilde{F}_{JX/Yq} &=&2\,\mbox{Re}\left\{ \tilde{F}_{X/Y}v_{Jq}\right\} 
\\
\delta \tilde{T}_{Jq} &=&2\,\mbox{Re}\left\{ \tilde{T}_{\psi }^{\left(
1\right) }v_{Jq}\right\} 
\end{eqnarray*}
$\tilde{F}_{X/Y},\tilde{T}_{\psi }$ being defined in Appendix C$.$
Accordingly, in the ideal setting of the cavity the forces and torques are
respectively proportional to longitudinal $\delta \tilde{\psi}_{J,cav}$ and
transverse $\delta \tilde{\alpha}_{Jq,cav}^{\prime \prime }$ fluctuations
through the stiffness coefficients $\tilde{F}_{\psi },\tilde{T}_{X/Y}$. A
slight deviation from it introduces forces and torques with a reverse
dependence on fluctuations, say $\delta \tilde{F}_{J}^{\left( 1\right)
},\delta \tilde{T}_{Jq}^{\left( 1\right) }$ depend respectively on $\delta 
\tilde{\alpha}_{Jq,cav\,}$ and $\delta \tilde{\psi}_{J,cav}.$ \ 

\section{Error signals \label{sec:Error-signals}}

The errors used for controlling the cavity are provided by Drever-Pound (DP)
and quadrant detector signals (QD). In the DP detection technique the
photodetector current $I\left( t\right) $, obtained from the light
transmitted and reflected by the input mirror, is mixed with a local
oscillator $\sim \sin \left( k\Lambda t+\varphi \right) $ with positive
odd integer k and low-pass filtered by an averaging procedure 
\begin{equation}
s^{DP}\left( t\right) =\int_{-\infty }^{t}K^{DP}\left( t-t^{\prime }\right)
\sin \left( k\Lambda t^{\prime }+\varphi \right) I\left( t^{\prime }\right)
dt^{\prime }  \label{DP integral signal}
\end{equation}
with the filter response $K^{DP}\left( t-t^{\prime }\right) $ extended to a
suitable interval much longer than $\left( k\Lambda \right) ^{-1}$, and
short compared to the time scale of the phase-quadrature fluctuations.
Tuning $%
\varphi $ around 0 $s^{DP}$ can be maximized for a misaligned cavity.

Putting $\delta 
\mathbf{\tilde{a}}_{p}^{SN}=\delta \mathbf{\hat{a}}^{SN}\left( \varpi
+p\Lambda \right) $, $s^{DP}$ is represented in the frequency domain by 
\begin{eqnarray}
\tilde{s}^{DP}/\tilde{K}^{DP} &=&\mathcal{E}^{2}\left( \bar{s}^{DP}\tilde{\mu%
}^{\ell }+\tilde{\mathfrak{s}}^{DP}\cdot \mathfrak{\delta \alpha }%
_{cav}+\delta \tilde{s}^{DP\;DEF}\right)   \nonumber \\
&&\hspace{1in}+\mathcal{E}\tilde{X}^{DP\;SN}  \label{DP-signal}
\end{eqnarray}
where 
\begin{eqnarray*}
\bar{s}^{DP} &=&\mathbf{v}_{1}^{\dagger }\cdot \mathbf{I}^{DP}\cdot \mathbf{v%
}_{1} \\
\tilde{\mathfrak{s}}^{DP} &=&\mathbf{v}_{1}^{\dagger }\cdot \left( %
\mathfrak{I}_{+}^{DP\,\,}-\mathfrak{I}_{-}^{DP\,\,}\right) \cdot \mathbf{v}%
_{1} \\
\delta \tilde{s}^{DP\;DEF} &=&\mathbf{v}_{1}^{\dagger }\cdot \left( \delta 
\mathbf{\tilde{I}}_{+}^{DP\,\,DEF\,\,}-\delta \mathbf{\tilde{I}}%
_{-}^{DP\,\,\,\,DEF\,}\right) \cdot \mathbf{v}_{1} \\
\tilde{X}^{DP\;SN} &=&\tilde{X}_{+}^{DP\;SN}-\tilde{X}_{-}^{DP\;SN}
\end{eqnarray*}
with $\mathbf{I}^{DP},\delta \mathbf{\tilde{I}}_{\pm }^{DP\,\,DEF\,\,},%
\tilde{\mathfrak{I}}_{\pm }^{DP\,\,\,},\tilde{X}_{\pm }^{DP\;SN}$ defined in
Appendix B.

\begin{figure}
\includegraphics[width=5cm]{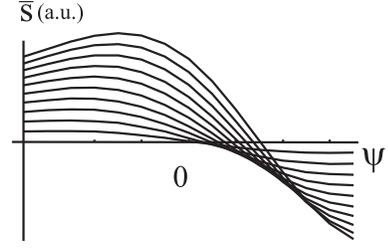}
\caption{
Drever-Pound static characteristics \={s} vs. $\psi $ for cavity
lengths $1\div 10$ cm and $\varphi =0$. The plots correspond to modulation
frequency $\Lambda =2c\lambda ^{\ell }/\left( \pi w^{2}\right) $, depth $%
M=0.1$, and jaw angle $\delta \theta _{z}=0.01$.
\label{Fig3}}
\end{figure}

In Figure \ref{Fig3} the static characteristic $\bar{s}^{DP}$ versus $\psi $ has
been plotted for a set of cavity lengths and modulations.

Figure 4 contains plots of the coefficients $s_{\psi }$ and $s_{Xy}$ vs.
cavity length for $\varphi =0,$ $\theta z=.1$ mrad and 7 detunings. They
show that as a consequence of the misalignment $s_{Xy}$ becomes comparable
to $s_{\psi },$ so that the D-P error signal contains contributions of the
torsional fluctuations around the vertical axis.

\begin{figure}
\includegraphics[width=6cm]{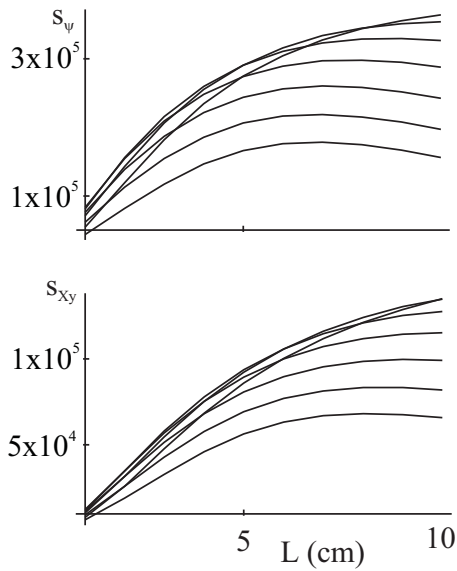}
\caption{
Coefficients $s_{\psi }$ and $s_{Xy}$ vs. cavity length for $%
\varphi =0,$ $\theta z=.1$ mrad and 7 detunings ranging in the interval $-.3%
\frac{\pi }{\mathcal{F}}\leq \psi \leq .3\frac{\pi }{\mathcal{F}}$.
\label{Fig4}}
\end{figure}

At low frequency $\delta \tilde{s}^{DP\;DEF}$ becomes proportional to the
thermal noises $\delta \tilde{\varsigma}^{L\;TH}.$

The quadrant detector used for stabilizing the angular oscillations provides
two error signals $s_{q}^{QD}\left( t\right) $ ($q=y,z$), proportional to
expressions similar to~(\ref{DP integral signal}) with the current $I$
replaced by $I_{q}=\mathbf{\beta }^{\dagger }\cdot \mathbf{Q}_{q}\cdot 
\mathbf{\beta }$ with 
\[
\mathbf{\beta }=e^{-ip\Lambda t}\mathbf{a}_{p}^{OUT}+\delta \mathbf{\hat{a}}%
^{OUT\;SN}\ \,,
\]
the matrix $\mathbf{Q}_{q}$ representing the function $sgn\left( \bar{q}%
\right) $. Then, $\tilde{s}_{q}^{QD}$ is given by an expression similar to (%
\ref{DP-signal}) with $\mathbf{G}_{p}^{OUT\dagger }$ replaced by $\mathbf{G}%
_{p}^{OUT\dagger }\cdot \mathbf{Q}_{q}$.

\subsection{\textbf{Small misalignment and mismatch}}

In the limit of small misalignment and mismatch the signal can be split into
zeroth- and first-order contributions 
\[
\tilde{s}^{DP}=\tilde{s}^{\left( 0\right) }+\delta \tilde{s}^{\left(
1\right) } 
\]
given respectively by (Eqs. (\ref{D-P expansion coefficients})) 
\begin{eqnarray}
\tilde{s}^{\left( 0\right) }/\tilde{K}^{DP}=\mathcal{E}^{2}\left( \bar{s}%
_{\psi }\tilde{\mu}^{\ell }+\tilde{s}_{\psi }\delta \tilde{\psi}%
_{1,cav}+\delta \tilde{s}^{\left( 0\right) \;DEF}\right)  \nonumber \\
\hspace{1in} +\mathcal{E}\tilde{X}^{SN\;\left( 0\right) }  \nonumber \\
\delta \tilde{s}^{\left( 1\right) }/\tilde{K}^{DP}=\mathcal{E}^{2}\left(
\delta \tilde{s}_{Xq}\delta \tilde{\alpha}_{1q,cav\,}^{\prime \prime
}+\delta \tilde{s}_{Yq}\delta \tilde{\alpha}_{1q,cav\,}^{\prime } \right. 
\nonumber \\
\hspace{1in} \left. +\delta \tilde{s}^{\left( 1\right) \;DEF}\right) +%
\mathcal{E}\delta \tilde{X}^{SN\;\left( 1\right) }
\end{eqnarray}
with $\delta \tilde{s}_{X/Yq}=2\mbox{Re}\left\{ \tilde{s}_{X/Y}v_{1q}\right\} $
and $\tilde{s}_{X/Y}$ defined in Appendix C$.$

For a perfectly aligned and matched cavity the \mbox{D-P} signal is
sensitive to the axial fluctuations $\delta \tilde{\psi}_{1,cav}$, mirror
deformation term $\delta \tilde{s}^{\left( 0\right) \;DEF}$ and shot noise $%
\tilde{X}^{SN\;\left( 0\right) }.$ In particular, a contribution $\delta 
\tilde{\psi}_{1,cav}$ depending on the mirror thermal noise $\delta \tilde{%
\varsigma}_{1,2}^{L}$ is added to the length fluctuation. A deviation from
alignment introduces in the error signal contributions proportional to the
transverse fluctuations.

\section{3D model\label{sec:3D-model}}

The deviations of each mirror from the reference position is described by
the displacements $\delta x,\delta \varepsilon _{y}$ and $\delta \varepsilon
_{z}$ of its vertex and the angular parameters $\delta \theta _{z}=-\delta
\Omega _{y}$ and $\delta \theta _{y}=\delta \Omega _{z}$. As said $\delta
\Omega _{q}$ describes a right-handed rotations around the axis ''$q$'', so
that $\delta \theta _{z}$ is a left-handed tilt and $\delta \theta _{y}$ a
right-handed torsion. These quantities fluctuate as a consequence of
suspension thermal fluctuations and mirror surface deformations. The
radiation pressure fluctuations are transferred to the mirrors
proportionally with the laser intensity. The cavity reacts by changes of
geometry which in turn changes the stored field and closes the loop of the
cavity-field system.

From a purely-mechanical point of view if the design is good (that is,
symmetric enough) the suspension masses are aligned along the vertical axis
z, perpendicular to the cavity axis x. In these conditions the torsion $%
\delta \theta _{y}$ and vertical $\delta \varepsilon _{z}$ degrees of
freedom are uncoupled. A coupling between longitudinal motion $\delta x$ and
tilt $\delta \theta _{z}$ is generally speaking unavoidable. This is true
also for the transverse displacement $\delta \varepsilon _{y}$ which is
coupled with the rotation around the optical axis. It goes without saying
that in a real situation it is very difficult to avoid more general cross
couplings.

Radiation pressure can increase or reduce existing couplings, and it can
also produce new ones. While $\delta \varepsilon _{q}$ is insensitive to
radiation pressure, $\delta \theta _{q}$ responds to the radiation torque.
For this reason when asymmetric optical modes are excited the rotations $%
\delta \theta _{q}$ modify the radiation pressure, and ultimately couple $%
\delta x$ and tilting, but also $\delta x$ and torsion.

Before proceeding further it is worth replacing the displacements $\delta
\varepsilon _{Jq}$ by $\delta \psi _{Jq}=2k^{\ell }\delta \varepsilon _{Jq}$%
, the angles $\delta \theta _{Jq}$ by $\delta \vartheta _{Jq}=\sqrt{2}%
k^{\ell }w_{J}\delta \theta _{Jq}$ and introducing a new five component
vector $\delta \mathbf{\psi }_{J}=\left( \delta \psi _{J},\delta \psi
_{Jq},\delta \vartheta _{Jq}\right) $ which forms with the cavity mode
amplitudes a system of correlated stochastic processes. It is usually a very
good approximation to model the suspension as a set of damped, independent
oscillators coupled to an heat bath. Each oscillator $J\lambda\ \hat{\jmath}$,
labelled by $\lambda$, specifying the prevalent character of the mode
(tilting, torsion, displacements,violin modes), and the mode index $\hat{%
\jmath}$, can be parameterized with its effective mass $M_{J\lambda \hat{%
\jmath}},$ pulsation $\varpi _{J\lambda \hat{\jmath}}$ and damping
coefficient $\gamma _{J\lambda \hat{\jmath}}$. For rotations $M_{J\lambda 
\hat{\jmath}}$ is replaced by the moment of inertia. These parameters are
related to the masses and stiffness constants of the system. The coordinates
of the mirror can be written as linear combinations of the oscillator's
coordinates $q_{n}$, and this means that each normal mode gives in principle
a contribution to the mirror's motion. By interacting with thermal baths
these modes undergo Brownian motions by influencing the electromagnetic
field, eventually coupling mechanical and radiation pressure fluctuations.

\subsection{Suspension Langevin system}

By linearizing the equation of motion of each mirror ($J$) the horizontal ($x
$ and $y$) and vertical ($z$) displacements $\delta \mathbf{\tilde{\psi}}_{J}
$, torsion $\delta \tilde{\vartheta}_{Jy}$, tilt $\delta \tilde{\vartheta}%
_{Jz}$ and rotation around the cavity axis $\delta \tilde{\vartheta}_{Jx}$
are expressed in terms of the amplitudes $\breve{A}_{J\lambda \hat{\jmath}}$
of the normal modes as 
\[
\delta \tilde{\psi}_{J\mu }=K_{J\mu \lambda \hat{\jmath}}\breve{A}_{J\lambda 
\hat{\jmath}}
\]
having indicated by $J\mu $ a generic degree of freedom and by $K_{J\mu
\lambda \hat{\jmath}}$ the coupling coefficient with the mode $J\lambda \hat{%
\jmath}$~\cite{Vicere}.

If the mirror vertex coincides with the center of mass of the suspension
payload, and the centers of the suspended masses are aligned along the
vertical $z$-axis, the suspension can be easily modeled by considering only
the couplings $\delta \psi _{J}-\delta \tilde{\vartheta}_{Jz}$ and $\delta
\psi _{Jy}-\delta \tilde{\vartheta}_{Jx}$, and assuming the vertical
oscillations independent of the other degrees of freedom. Being the
amplitudes of the cavity modes independent of the rotations $\delta \tilde{%
\vartheta}_{Jx}$, the suspended cavity is described by the collection $%
\mathbf{\delta \tilde{\psi}}_{J}$ of five fluctuating quantities, depending
linearly on radiation pressure-torques, thermal noise, \mbox{D-P} and
quadrant detector error signals, 
\begin{eqnarray}
\delta \tilde{\psi}_{J} &=&\tilde{\chi}_{J\psi }\,8\frak{R}_{J}\left( 
\mathcal{E}^{2}\delta \tilde{F}_{J}+\mathcal{E}\tilde{X}_{J}^{SN}\right)  
\nonumber \\
&&+\tilde{\chi}_{J\psi \theta z}8\frak{R}_{J}\left( \mathcal{E}^{2}\delta 
\tilde{T}_{Jz}+\mathcal{E}\tilde{X}_{J\theta z}^{SN}\right)   \nonumber \\
&&+\tilde{\mathcal{X}}_{J\psi }^{TH}+\delta _{J1}H^{DP}\tilde{s}^{DP} 
\nonumber \\
\delta \tilde{\psi}_{Jy/z} &=&\tilde{\mathcal{X}}_{Jy/z}^{TH}  \nonumber \\
\delta \tilde{\vartheta}_{Jz} &=&\tilde{\chi}_{J\theta z\psi }8\frak{R}%
_{J}\left( \mathcal{E}^{2}\delta \tilde{F}_{J}+\mathcal{E}\tilde{X}%
_{J}^{SN}\right)   \nonumber \\
&&+\tilde{\chi}_{J\theta z}8\frak{R}_{J}\left( \mathcal{E}^{2}\delta \tilde{T%
}_{Jz}+\mathcal{E}\tilde{X}_{J\theta z}^{SN}\right)   \nonumber \\
&&+\tilde{\mathcal{X}}_{J\theta z}^{TH}+\delta _{J1}H_{z}^{QD}\tilde{s}%
_{z}^{QD}  \nonumber \\
\delta \tilde{\vartheta}_{Jy} &=&\tilde{\chi}_{J\theta y}8\frak{R}_{J}\left( 
\mathcal{E}^{2}\delta \tilde{T}_{Jy}+\mathcal{E}\tilde{X}_{J\theta
y}^{SN}\right)   \nonumber \\
&&+\tilde{\mathcal{X}}_{J\theta y}^{TH}+\delta _{J1}H_{y}^{QD}\tilde{s}%
_{y}^{QD}  \label{Langevin system}
\end{eqnarray}
In case the mirror vertex and/or the centers of the suspension wire
clampings are displaced from the respective mass centers, the vertical
fluctuations are coupled to the other ones.

The effect of the servo systems acting on the longitudinal and angular
mirror displacements have been included by indicating by $\tilde{H}^{DP}$
and $\tilde{H}_{q}^{QD}$ the respective transfer functions.

For the mirror vibrations a Langevin equation for each mode must be
considered since their profiles are different (Eq.~(\ref{mirror mirror
forces})), 
\begin{equation}
\delta \tilde{\varsigma}_{Js}=8\tilde{\chi}_{Js}\left( \mathcal{E}^{2}\delta 
\tilde{F}_{Js,cav}^{DEF}+\mathcal{E}\widetilde{X}_{Js}^{DEF\;SN}\right) +%
\tilde{\mathcal{X}}_{Js}^{DEF\;TH}  \label{Langevin mirror modes}
\end{equation}

Expressing $\delta \tilde{F}_{J},\delta \tilde{T}_{Jq}$ in terms of
displacements and rotations by introducing the stiffness coefficients, and
doing the same for the error signals $\tilde{s}^{DP},\tilde{s}_{q}^{QD}$ the
above system can be reduced to an equivalent one relating the fluctuating
displacement + rotations to the thermal noise and shot noise sources.

The axial displacement $\delta \tilde{\psi}_{J}$ and tilting $\delta \tilde{%
\vartheta}_{Jz}$ respond to the axial force $\mathcal{E}^{2}\delta \tilde{F}%
_{J}+\mathcal{E}\tilde{X}_{J}^{SN}$ and torque $\left( \mathcal{E}^{2}\delta 
\tilde{T}_{Jz}+\mathcal{E}\tilde{X}_{J\theta z}^{SN}\right) \hat{y}$
generated by the laser beam and the shot noise. By the way they include the
contributions of the mirror thermal noise. On the other hand, $\delta \tilde{%
\vartheta}_{Jy}$ responds to the torque $\mathcal{E}^{2}\delta \tilde{T}%
_{Jy}+\mathcal{E}\tilde{X}_{J\theta y}^{SN}$. The links between
force-torques and $\mathfrak{\delta}\tilde{\mathfrak{\psi}}_{J}$ are
represented by the susceptibilities $\tilde{\chi}_{J\mu \nu }$.

The terms proportional to $\mathcal{E}^{2}$ and $\mathcal{E}$ describe the
response of the system to radiation pressure. Their presence indicates that
a motion of the mirrors produces not only a phase change but also an
intensity change providing a spring action.

In writing Eq.~(\ref{Langevin system}) the interaction with the mirror noise
was approximated with Eq.~(\ref{low frequency deformation}) while in Eq.~(%
\ref{Langevin mirror modes}) the effects of the suspension fluctuations were
ignored. Loosely speaking the two systems refer respectively to the low and
high frequency regions. In the former one the suspensions are mutually
coupled by radiative forces represented while the mirror vibrations generate
a global thermal noise hiding the single mode contributions. In the latter
one the suspensions appear frozen and the mirror modes are mutually coupled
by radiative forces represented by $\delta \tilde{F}_{Js,cav}^{DEF}$.

The solutions of the homogeneous system (\ref{Langevin system}) represent,
in absence of feedback forces, free mechanical oscillations of the suspended
cavity, stable or unstable in accordance with the sign of the imaginary part
of the oscillation frequency \cite{spring}.

For a more detailed analysis (\ref{Langevin system}) and (\ref{Langevin
mirror modes}) should be mirrored by the system relative to the quantities $%
\mathbf{\delta }\tilde{\mathbf{\psi }}_{J}^{Y}$, $\mathfrak{\delta }\tilde{%
\boldsymbol{\varsigma}}_{J}^{Y}$ conjugate of $\mathfrak{\delta }\tilde{%
\mathbf{\psi }}_{J}$, $\left\{ \delta \varsigma _{Js}\right\} ,$ which can
be obtained from the above one by replacing $\tilde{\chi}_{J\psi /\theta q/s}
$ by $\tilde{\chi}_{J\psi /\theta q/s}^{Y}$ (Eq. \ref{susceptivity Y}) and $%
\widetilde{X}_{J\psi /\theta q}^{TH},\widetilde{X}_{Js}^{DEF\;TH}$ by $%
\tilde{Y}_{J\psi /\theta q}^{TH},\tilde{Y}_{Js}^{DEF\;TH}$ (Eq. \ref{thermal
noise Y}) in the random force expressions.

\subsection{Susceptibilities}

The susceptibility $\tilde{\chi}_{J\mu \nu }$ describes the action on the
coordinate $\mu $ of the force/torque acting on $\nu ,$ 
\[
\tilde{\chi}_{J\mu \nu }=K_{J\mu \lambda \hat{\jmath}}\,K_{J\nu \lambda \hat{%
\jmath}}\,\tilde{\chi}_{J\lambda \hat{\jmath}}
\]
with $\tilde{\chi}_{J\lambda \hat{\jmath}}$ the susceptibility of the mode $%
J\lambda \hat{\jmath}$ of frequency $\varpi _{J\lambda \hat{\jmath}}$ and
damping coefficient $\gamma _{J\lambda \hat{\jmath}}$ 
\begin{equation}
\tilde{\chi}_{J\lambda \hat{\jmath}}=\frac{\varpi _{J\lambda \hat{\jmath}%
}\,\eta _{J\lambda \hat{\jmath}}^{LD\;2}}{\varpi _{J\lambda \hat{\jmath}%
}^{2}-\varpi ^{2}-i\varpi \,\gamma _{J\lambda \hat{\jmath}}}
\label{susceptivities}
\end{equation}
and $K_{J\mu \lambda \hat{\jmath}}$, $K_{J\nu \lambda \hat{\jmath}}$ the
coupling coefficients with $\mu $ and $\nu $ mirror coordinates, while the
adimensional Lamb-Dicke factor 
\begin{equation}
\eta _{J\lambda \hat{\jmath}}^{LD}=k^{\ell }\sqrt{\frac{\hbar }{2M_{J\lambda 
\hat{\jmath}}\,\varpi _{J\lambda \hat{\jmath}}}}  \label{Lamb-Dicke factor}
\end{equation}
depends on the mode mass $M_{J\lambda \hat{\jmath}}=M_{Ji}K_{Ji\lambda \hat{%
\jmath}}^{2}$ (the subfix $i$ identifies the $i$--th mass of the
suspension). For rotations $M_{J\lambda \hat{\jmath}}$ is replaced by $%
I_{J\lambda \hat{\jmath}}/w_{J}^{2}$ with $I_{J\lambda \hat{\jmath}}$ the
moment of inertia. Some authors use the so-called optomechanical coupling
constants $G_{J\lambda \hat{\jmath}}=2\sqrt{2}\eta _{J\lambda \hat{\jmath}%
}^{LD}/\tau $ \cite{Mancini}.

The mechanical susceptibility $\tilde{\chi}_{Js}$ is similar to (\ref
{susceptivities}) while the mass appearing in the Lamb-Dicke factor varies
for the different modes, as reported in \cite{Hadjar}.

\subsection{Thermal contributions}

Assuming suspension masses at the same temperature T, each mode is
characterized by a thermal source\ (see Appendix D) 
\begin{equation}
\tilde{X}_{J\lambda \hat{\jmath}}^{TH}=\sqrt{\frac{4k_{B}T}{\hbar \varpi
_{J\lambda \hat{\jmath}}}}\tilde{\xi}_{J\lambda \hat{\jmath}}-i\frac{\varpi
+i\gamma _{J\lambda \hat{\jmath}}}{\varpi _{J\lambda \hat{\jmath}}}\sqrt{%
\frac{\hbar \varpi _{J\lambda \hat{\jmath}}}{3k_{B}T}}\tilde{\eta}_{J\lambda 
\hat{\jmath}}  \label{thermal noise}
\end{equation}
with $\tilde{\eta},\tilde{\xi}$ delta correlated random forces introduced by
Diosi \cite{Jacobs} in order to remove some inconsistencies of the classical
Langevin equation.

A Y-version of (\ref{Langevin system}) can be easily obtained for the
Y-quadratures corresponding to the above ones by replacing $\tilde{\chi}%
_{J\mu \lambda \hat{\jmath}}$ by 
\begin{equation}
\tilde{\chi}_{J\lambda \hat{\jmath}}^{Y}=i\frac{\varpi }{\varpi _{J\lambda 
\hat{\jmath}}}\tilde{\chi}_{J\lambda \hat{\jmath}}  \label{susceptivity Y}
\end{equation}
and $\tilde{X}_{J\lambda \hat{\jmath}}^{TH}$ by 
\begin{equation}
\tilde{Y}_{J\lambda \hat{\jmath}}^{TH}=\sqrt{\frac{4k_{B}T}{\hbar \varpi
_{J\lambda \hat{\jmath}}}}\tilde{\xi}_{J\lambda \hat{\jmath}}-i\frac{\varpi
_{J\lambda \hat{\jmath}}}{\varpi }\sqrt{\frac{\hbar \varpi _{J\lambda \hat{%
\jmath}}}{3k_{B}T}}\tilde{\eta}_{J\lambda \hat{\jmath}}
\label{thermal noise Y}
\end{equation}

The terms proportional to $\tilde{\eta}_{J\hat{\jmath}}$ in Eqs. (\ref
{thermal noise}) and (\ref{thermal noise Y}) can be generally neglected
except when the temperature is rather low and the oscillation frequencies
very high, a situation met only in some mirror modes.

$\tilde{\eta}_{J\lambda \hat{\jmath}}$ disappears in the simple Brownian
motion model while in Ref. \cite{Giovannetti} $\tilde{\eta}_{J\lambda \hat{%
\jmath}}$ has been dropped and $\sqrt{\frac{4k_{B}T}{\hbar \varpi _{J\lambda 
\hat{\jmath}}}}\tilde{\xi}_{J\lambda \hat{\jmath}}$ replaced by a new delta
correlated random noise source $\tilde{Q}_{J\lambda \hat{\jmath}}.$

The thermal sources $\tilde{\mathcal{X}}_{J\mu }^{TH}$ are superpositions 
\[
\tilde{\mathcal{X}}_{J\mu }^{TH}=K_{J\mu \lambda \hat{\jmath}}\,\tilde{\chi}%
_{J\lambda \hat{\jmath}}^{TH}\,\tilde{X}_{J\lambda \hat{\jmath}}^{TH}
\]
of the $\tilde{X}_{J\lambda \hat{\jmath}}^{TH}$ weighted by the thermal
susceptivities 
\begin{equation}
\tilde{\chi}_{J\lambda \hat{\jmath}}^{TH}=\kappa _{J\lambda \hat{\jmath}}\,%
\tilde{\chi}_{J\lambda \hat{\jmath}}  \label{thermal susceptivities}
\end{equation}
with  $\kappa _{J\lambda \hat{\jmath}}=2\sqrt{\gamma%
_{J\lambda \hat{\jmath}}}/\eta _{J\lambda \hat{\jmath}}^{LD}.$

The terms of (\ref{Langevin system}) contain contributions proportional to
the fluctuating quantities $\delta \tilde{\varsigma}_{J}^{L\;TH}$ \cite
{Levin} 
\begin{equation}
\delta \tilde{\varsigma}_{J}^{L\;TH}=\sqrt{\frac{4k_{B}T}{\hbar \varpi }}%
\sqrt{2\hbar k^{\ell }c_{P}\phi _{J}}\tilde{\varsigma}
\label{spectrum mirror noise}
\end{equation}
with $\phi $ the loss angle, $\tilde{\varsigma}$ a delta correlated
random force and $c_{P}$ depending on the illumination profile 
\begin{equation}
P\left( \vec{r}\right) =P_{\lambda _{y},\lambda _{z}}e^{-\frac{r^{2}}{2w^{2}}%
}u_{\lambda _{y},\lambda _{z}}\left( \vec{r}\right) 
\label{illumination expansion}
\end{equation}
For $P\left( \mathbf{\vec{r}}\right) $ differing notably from the Gaussian
one the deformed profile of the mirror $\delta u_{1,2}^{DEF}$ \ can be
expressed, neglecting the finite size of the mirrors, by a suitable
combination of derivatives of the deformation $\delta u_{G}^{DEF}\left( \vec{%
r}\right) $ relative to a Gaussian distribution 
\[
\delta u^{DEF}\left( \vec{r}\right) =\sum_{\lambda _{y},\lambda
_{z}}P_{\lambda _{y},\lambda _{z}}\left( -w\right) ^{\lambda _{y}+\lambda
_{z}}\frac{\partial ^{\lambda _{y}}}{\partial y^{\lambda _{y}}}\frac{%
\partial ^{\lambda _{z}}}{\partial z^{\lambda _{z}}}\delta u_{G}^{DEF}\left( 
\vec{r}\right) 
\]
For a Gaussian illumination $c_{P}$ \ takes the form 
\[
c_{G}=\frac{1-\sigma ^{2}}{\sqrt{2\pi }Ew_{J}}
\]
with $w_{J}$ the spot-size and $E$,$\sigma $ the Young's modulus and Poisson
ratio respectively. For a generic illumination $c_{P}$ can be expressed as $%
c_{P}=f_{P}\,c_{G}$ with 
\begin{equation}
f_{P}=\sum_{\lambda \lambda ^{\prime }}\left( -1\right) ^{\lambda
_{y}^{\prime }+\lambda _{z}^{\prime }}\frac{P_{\lambda _{y},\lambda
_{z}}P_{\lambda _{y}^{\prime },\lambda _{z}^{\prime }}}{P_{00}^{2}}%
f_{\lambda _{y}+\lambda _{y}^{\prime },\lambda _{z}+\lambda _{z}^{\prime }}
\label{spectrum factor}
\end{equation}
$f_{\alpha \beta }$ being the $\alpha \beta $ coefficient of the expansion
of $\delta u_{G}\left( \vec{r}\right) e^{-\frac{r^{2}}{2w^{2}}}$ in modes $%
u_{\lambda _{y},\lambda _{z}}\left( \vec{r}\right) $.

\section{The suspended cavity as a bipartite system\label%
{sec:bipartite system}}

When the frequency is in proximity of two close resonances of the mirror
$1$ and $2$ modes, the system behaves as a quantum mechanical
bipartite system described by Gaussian continuous variables.\ These systems
can form EPR states characterized by their covariance matrix $\mathbf{\sigma 
}$ which can be used for evaluating the entanglment of the state and its
content of quantum information.

The difference between the e.m. fields used in quantum optics and the
present mechanical system concerns the sources of the respective states. The
e.m. fields are produced by the e.m. vacuum noise entering through the
mirrors of a cavity containing a nonlinear crystal. In the present case
thermal and shot noises act as sources. Accordingly, the covariance matrix $%
\mathbf{\sigma }$ can be split into thermal $\mathbf{\sigma }^{TH}$ and shot
noise $\left( 8\mathcal{E}\right) ^{2}\mathbf{\sigma }^{SN}$ contributions
obtained by separating $\delta \tilde{\varsigma}_{J}$ into $\delta \tilde{%
\varsigma}_{J}=8\mathcal{E}\delta \tilde{\varsigma}_{J}^{SN}+\delta \tilde{%
\varsigma}_{J}^{TH}$ satisfying the Langevin system (\ref{Langevin mirror
modes})

\begin{equation}
\left[ 
\begin{array}{c}
\delta \tilde{\varsigma}_{1}^{SN/TH} \\ 
\delta \tilde{\varsigma}_{2}^{SN/TH}
\end{array}
\right] =\frac{1}{D}\left[ 
\begin{array}{cc}
\tilde{\mathcal{P}}_{22} & -\tilde{\mathcal{P}}_{12} \\ 
-\tilde{\mathcal{P}}_{21} & \tilde{\mathcal{P}}_{11}
\end{array}
\right] \cdot \left[ 
\begin{array}{c}
\chi _{1}\tilde{X}_{1}^{DEF\;SN/TH} \\ 
\chi _{2}\tilde{X}_{2}^{DEF\;SN/TH}
\end{array}
\right]   \label{mirror quadratures}
\end{equation}
with $\tilde{\mathcal{P}}_{JJ^{\prime }}$ factors representing the radiation
pressure effects 
\begin{eqnarray*}
\tilde{\mathcal{P}}_{JJ} &=&1-8e^{i\varpi \tau }\mathcal{E}^{2}\,\frak{R}%
_{J}\,\tilde{\chi}_{J}\tilde{F}_{JJ}^{DEF} \\
\tilde{\mathcal{P}}_{J\bar{J}} &=&8e^{i\varpi \tau /2}\mathcal{E}^{2}\,\frak{%
R}_{J}\,\tilde{\chi}_{J}\tilde{F}_{J\bar{J}}^{DEF}
\end{eqnarray*}
and their product $\tilde{D}=\tilde{\mathcal{P}}_{11}\tilde{\mathcal{P}}%
_{22}-\tilde{\mathcal{P}}_{12}\tilde{\mathcal{P}}_{21}$. 
An analogous system holds for$\mathcal{\ }\delta \tilde{\mathbf{\varsigma }}%
_{J}^{Y\;SN}$ with $\tilde{\chi}_{J}$ replaced by $\tilde{\chi}_{J}^{Y}$.

The output field contains a component (Eqs. (\ref{approximate Green},\ref
{aout})) 
\[
\delta \mathbf{\tilde{a}}^{OUT}\propto \left( e^{i\varpi \tau }\tilde{%
\mathbf{Z}}_{1}\delta \tilde{\varsigma}_{1}+e^{i\varpi \tau /2}\tilde{%
\mathbf{Z}}_{2}\delta \tilde{\varsigma}_{2}\right) \mathbf{\cdot v}_{1}
\]
proportional to $\delta \tilde{\varsigma}_{1,2}$ through the matrices $%
\tilde{\mathbf{Z}}_{J}=\tilde{\mathbf{G}\cdot \Phi \cdot \varsigma }%
_{J}\cdot \tilde{\mathbf{G}}$ and a shot noise $\mathbf{\hat{G}}^{OUT}%
\mathbf{\cdot }\delta \mathbf{\hat{a}}_{1}^{SN}+\frac{t_{2}}{t_{1}}\mathbf{%
\hat{G}\cdot }\delta \mathbf{\hat{a}}_{2}^{SN}$ \ term. Hence, depending $%
\delta \tilde{\varsigma}_{1,2}^{SN}$ linearly on the quadratures $\tilde{X}%
_{1,2}^{DEF\;SN}$ the output exhibits some degree of squeezing., a feature
exploited by several groups in the context of gravitational antennas of the
next generation \cite{Heidmann}. The dependence of the efficiency of the
ponderomotive squeezing on the mirror deformation profiles (matrices $\tilde{%
\mathbf{Z}}_{J}$) and residual misalignment/mismatch can be easily analyzed
by means of Eqs. (\ref{mirror quadratures}) and the correlations of Apps. E
and F.

The complex dynamics of cavity field and ponderomotive effects\ may lead to
the creation of quantum entangled states of the two mirror modes, as shown
by Mancini et al. (\cite{Mancini} and references therein included). These
authors have proposed a measure $\Bbb{E}\left( \varpi \right) $ of the
entanglement degree (the smaller $\Bbb{E}\left( \varpi \right) <1$ the
larger the entanglement) based on a combination of the elements of the
covariance matrix, 
\begin{equation}
\Bbb{E}\left( \varpi \right) =\frac{\overline{|\delta \tilde{\mathbf{%
\varsigma }}_{1}+\delta \tilde{\mathbf{\varsigma }}_{2}|^{2}}\,\overline{%
|\delta \tilde{\mathbf{\varsigma }}_{1}^{Y}-\delta \tilde{\mathbf{\varsigma }%
}_{2}^{Y}|^{2}}}{\left| \overline{\left[ \delta \tilde{\mathbf{\varsigma }}%
_{1},\delta \tilde{\mathbf{\varsigma }}_{1}^{Y}\right] }\right| ^{2}}
\label{entanglment}
\end{equation}

Splitting the quadratures into shot noise and thermal contributions, taking
into account the many modes of the cavity and the shapes of the mirror
mechanical modes, and scaling the ratio terms by keeping constant $\Bbb{E}%
\left( \varpi \right) $, yield for the thermal and the shot noise
contributions 
\begin{eqnarray}
\overline{\left| \delta \tilde{\mathbf{\varsigma }}_{1}^{TH}+\delta \tilde{%
\mathbf{\varsigma }}_{2}^{TH}\right| ^{2}} &=&\left| \chi _{J}^{TH}\right|
^{2}\tilde{C}_{J}^{TH\;X\left( +\right) }
\label{spectral densities} \\
-\frac{i}{2}\overline{\left[ \delta \tilde{\mathbf{\varsigma }}%
_{1}^{TH},\delta \tilde{\mathbf{\varsigma }}_{1}^{TH\,Y}\right] } &=&\left( 
\frac{\varpi }{\varpi _{J}}\right) ^{2}\left| \alpha _{J}\,\chi
_{J}^{TH}\right| ^{2}  \nonumber \\
\overline{\left| \delta \tilde{\mathbf{\varsigma }}_{1}^{SN}+\delta \tilde{%
\mathbf{\varsigma }}_{2}^{SN}\right| ^{2}} &=&\chi _{J}\,\chi _{J^{\prime
}}^{\ast }\tilde{C}_{JJ^{\prime }}^{SN\;X\left( +\right) }  \nonumber \\
\overline{\left[ \delta \tilde{\mathbf{\varsigma }}_{1}^{SN},\delta \tilde{%
\mathbf{\varsigma }}_{1}^{SN\,Y}\right] } &=&\frac{\varpi \left( \varpi
_{J}+\varpi _{J^{\prime }}\right) }{2\varpi _{J}\varpi _{J^{\prime }}}\tilde{%
\chi}_{J}\,\tilde{\chi}_{J^{\prime }}^{\ast }\,\alpha _{J}\,\alpha _{\prime
}^{\ast }\,\tilde{C}_{JJ^{\prime }}^{SN\;Y} \nonumber
\end{eqnarray}
with $\chi _{J}^{TH}$ defined in (\ref{thermal susceptivities}), while$%
\overline{\left| \delta \tilde{\mathbf{\varsigma }}_{1}^{Y\;TH}-\delta 
\tilde{\mathbf{\varsigma }}_{2}^{Y\;TH}\right| ^{2}}$ and $\overline{\left|
\delta \tilde{\mathbf{\varsigma }}_{1}^{Y\;SN}-\delta \tilde{\mathbf{%
\varsigma }}_{2}^{Y\;SN}\right| ^{2}}$ are similar to (\ref{spectral
densities})--a and --c with $\chi _{J}$, $\tilde{C}_{J}^{TH\;X\left(
+\right) }$ and $\tilde{C}_{JJ^{\prime }}^{SN\;X\left( +\right) }$ replaced
respectively by $\chi _{J}^{Y}$, $\tilde{C}_{J}^{TH\;Y\left( -\right) }$ and 
$\frac{\varpi ^{2}}{\varpi _{J}\varpi _{J^{\prime }}}\tilde{C}_{JJ^{\prime
}}^{SN\;X\left( -\right) }$. On the other hand, 
\[
\left( \alpha _{1},\alpha _{2}\right) =\left( \tilde{\mathcal{P}}_{22},%
\tilde{\mathcal{P}}_{12}\right) \left| \tilde{\mathcal{P}}_{1}^{\left(
+\right) }\tilde{\mathcal{P}}_{2}^{\left( +\right) }\tilde{\mathcal{P}}%
_{1}^{\left( -\right) }\tilde{\mathcal{P}}_{2}^{\left( -\right) }\right|
^{-1/2}\,,
\]
$\tilde{\mathcal{P}}_{J}^{\left( \pm \right) }=\tilde{\mathcal{P}}_{JJ}\pm 
\tilde{\mathcal{P}}_{J\bar{J}}$ and 
\begin{eqnarray*}
\tilde{C}_{J}^{TH\;X/Y\left( \pm \right) } &=&\frac{\mbox{Re}\left\{ \tilde{C%
}_{J}^{XX/YY\;TH}\right\} }{2\left| \tilde{\mathcal{P}}_{J}^{\left( \pm
\right) }\right| ^{2}} \\
\tilde{C}_{JJ^{\prime }}^{SN\;X\left( \pm \right) } &=&\frac{\mbox{Re}%
\left\{ \tilde{C}_{JJ^{\prime }}^{SN}\right\} }{2\left| \tilde{\mathcal{P}}%
_{J}^{\left( \pm \right) \ast }\tilde{\mathcal{P}}_{J^{\prime }}^{\left( \pm
\right) }\right| } \\
\tilde{C}_{JJ^{\prime }}^{SN\;Y} &=&\frac{\mbox{Im}\left\{ \tilde{C}%
_{JJ^{\prime }}^{SN}\right\} }{2\sqrt{\left| \tilde{\mathcal{P}}_{1}^{\left(
+\right) }\tilde{\mathcal{P}}_{2}^{\left( +\right) }\tilde{\mathcal{P}}%
_{1}^{\left( -\right) }\tilde{\mathcal{P}}_{2}^{\left( -\right) }\right| }}
\end{eqnarray*}
with $\tilde{C}_{JsJ^{\prime }s^{\prime }}^{SN}$ given by Eq.~(\ref{shot
noise correlation operator}). In App. E thermal noise correlations for the
Lindblad--Diosi and the Giovanetti--Vitali MEs are explicitly given.

\section{\textbf{Conclusions}}

A suspended cavity illuminated by a laser beam has been described as the
mechanical response $\delta \mathbf{\psi }_{J}$ of each mirror of a linear
system to radiative, thermal and shot noise forces. These perturbations have
been linked to the mechanical responses by means of susceptibility
coefficients. The model includes the mirror vibrations described by a set of
mode amplitudes $\mathbf{\delta \varsigma}$, together with their shapes $%
\mathbf{\varsigma}$.

The radiative pressure forces and torques have been linearized with respect
to $\delta \mathbf{\psi }_{J}$ and $\mathbf{\delta \varsigma ,}$ by
obtaining sets of stiffness coefficients for the suspension ($\mathfrak{F}$
and $\mathfrak{T}$) and for the mirror modes ($F_{JsJ^{\prime }s^{\prime
}}^{DEF}$). Accordingly the radiative forces have been expressed as products
of susceptibility coefficients, laser intensity transmitted to the cavity ($%
\mathcal{E}^{2}$), stiffness coefficients, and mechanical mode amplitudes.
So doing $\delta \mathbf{\psi }_{J}$ and $\mathbf{\delta \varsigma }$ have
been linked directly to the thermal contributions, modelled by Lindblad and
non-Lindblad master equations, and to the shot noise forces. The mirror
thermal noise has been expressed in the low frequency limit by the Levin's
formula. A corrective factor, for taking into account deviations of the
cavity field from the fundamental mode, has been introduced.

The Drever-Pound and quadrant detector signals used for stabilizing
respectively longitudinally and angularly the cavity, have been expressed in
a form suitable to study the mutual coupling of these degrees of freedom in
case of misalignment.

Emphasis has been put on the description of missalignment and mismatch of
the input laser beam. To this end a vector approach has been adopted: the
state of the input beam and the amplitudes of the excited cavity modes have
been represented by vectors ($\mathbf{v}$ and $\mathbf{a}$, respectively)
and all the contributions to the cavity dynamic by a set of matrices. In
this way, all the relevant quantities are given in form of algebraic
products.

In particular, the optically-induced stiffness coefficients relative to the
suspension modes have been expressed in the form $\mathbf{v}^{\dagger }%
\mathbf{\cdot \mathfrak{F}\cdot v,v}^{\dagger }\mathbf{\cdot \mathfrak{T}%
\cdot v}$ with $\mathbf{\mathfrak{F,T}}$ matrices. It has been shown
numerically that these coefficients may become very large in misaligned
cavities close to unstable configurations.

The finite cavity round-trip time has been included in the model by
introducing a delay operator. Consequently the cavity has been represented
in the frequency domain by frequency dependent matrices containing stiffness
coefficients.

The reported model simplifies notably in proximity of mechanical resonances.
In particular the covariance matrix $\mathbf{\sigma }$ of two close in
frequency vibrational modes has been expressed in terms of the stiffness
coefficients and used for evaluating the system entanglement .This matrix
also controls the squeezing degree of the output field.

The numerical examples refer to almost concentric cavities of length varying
between $1$~cm and $10$~cm, spot size.$2$~cm and misalignment of $\ .1$~mrad.

\appendix

\section{Force, torques and stiffness operators}

\begin{eqnarray}
\mathbf{\bar{F}}_{0} &=&J_{p}^{2}\mathbf{G}_{p}^{\dagger }\cdot \mathbf{G}%
_{p}  \nonumber \\
\mathbf{\bar{T}}_{0,q} &=&J_{p}^{2}\mathbf{G}_{p}^{\dagger }\cdot \mathbf{X}%
_{q}\cdot \mathbf{G}_{p}  \label{F0T0}
\end{eqnarray}
Next, the stiffness operators $\tilde{\mathfrak{F}},\tilde{\mathfrak{T}}_{q}$
are given by 
\begin{eqnarray}
\tilde{\mathfrak{F}} &=&2J_{p}^{2}\Im \left\{ \tilde{\mathfrak{F}}%
_{p}\right\}  \nonumber \\
\tilde{\mathfrak{T}}_{q} &=&2J_{p}^{2}\Im \left\{ \tilde{\mathfrak{T}}%
_{qp}\right\}  \label{stiffness matrices}
\end{eqnarray}
with 
\begin{eqnarray}
\tilde{\mathfrak{F}}_{p} &=&e^{-i\psi }R_{p}\mathbf{G}_{p}^{\dagger }\cdot 
\tilde{\mathbf{G}}_{p}\cdot \mathbf{\Phi }\cdot \mathfrak{X}\cdot \mathbf{G}%
_{p}  \nonumber \\
\tilde{\mathfrak{T}}_{qp} &=&e^{-i\psi }R_{p}\mathbf{G}_{p}^{\dagger }\cdot 
\mathbf{X}_{q}\cdot \tilde{\mathbf{G}}_{p}\cdot \mathbf{\Phi }\cdot %
\mathfrak{X}\cdot \mathbf{G}_{p}  \label{stiffness terms}
\end{eqnarray}
while 
\begin{eqnarray}
\delta \tilde{\mathbf{F}}_{J,cav}^{DEF} &=&2J_{p}^{2}\Im \left\{ \delta 
\tilde{\mathbf{F}}_{Jp,cav}^{DEF}\right\}  \nonumber \\
\delta \tilde{\mathbf{T}}_{Jq,cav}^{DEF} &=&2J_{p}^{2}\Im \left\{ \delta 
\tilde{\mathbf{T}}_{Jqp,cav}^{DEF}\right\}  \label{FTDEF}
\end{eqnarray}
with 
\begin{eqnarray}
\delta \tilde{\mathbf{F}}_{Jp,cav}^{DEF} &=&e^{-i\psi }R_{p}\mathbf{G}%
_{p}^{\dagger }\cdot \tilde{\mathbf{G}}_{p}\cdot \mathbf{\Phi \cdot \delta} 
\tilde{\mathbf{\varsigma}}_{J,cav}^{L}\cdot \mathbf{G}_{p}  \nonumber \\
\delta \tilde{\mathbf{T}}_{Jqp,cav}^{DEF} &=&e^{-i\psi }R_{p}\mathbf{G}%
_{p}^{\dagger }\cdot \mathbf{X}_{q}\cdot \tilde{\mathbf{G}}_{p}\cdot \mathbf{%
\Phi \cdot \delta} \tilde{\mathbf{\varsigma}}_{J}^{L}\cdot \mathbf{G}_{p}
\label{FTDEFM}
\end{eqnarray}
where $\mathbf{\delta} \tilde{\mathbf{\varsigma}}_{J,cav}^{L}$ takes the
Levin's form, 
\[
\mathbf{\delta} \tilde{\mathbf{\varsigma}}_{J,cav}^{L}=e^{i\varpi \tau }%
\mathbf{\varsigma }_{J}^{L}\delta \tilde{\varsigma}_{J}^{L}+e^{i\varpi \tau
/2}\mathbf{\varsigma }_{\bar{J}}^{L}\delta \tilde{\varsigma}_{\bar{J}}^{L} 
\]

Finally the action of the modes $J^{\prime }s^{\prime }$ on the Js one is
represented by the ensemble of matrices 
\begin{equation}
\tilde{\mathbf{F}}_{JsJ^{\prime }s^{\prime }}^{DEF}=2J_{p}^{2}\Im \left\{ 
\tilde{\mathbf{F}}_{pJsJ^{\prime }s^{\prime }}^{DEF}\right\}
\label{stiffnes mirror modes}
\end{equation}
with

\begin{equation}
\tilde{\mathbf{F}}_{pJsJ^{\prime }s^{\prime }}^{DEF}=e^{-i\psi }R_{p}\mathbf{%
G}_{p}^{\dagger }\cdot \mathbf{\varsigma }_{Js}\cdot \tilde{\mathbf{G}}%
_{p}\cdot \mathbf{\Phi \cdot \varsigma }_{J^{\prime }s^{\prime }}\cdot 
\mathbf{G}_{p}  \label{deformation matrix}
\end{equation}

\section{\label{sec:Appendix-B}Drever-Pound signal}

\begin{eqnarray}
&&\mathbf{I}^{DP}=2J_{p-k}J_{p}\Im \left\{ e^{i\varphi }\mathbf{G}%
_{p}^{OUT\;\dagger }\cdot \mathbf{G}_{p-k}^{OUT}\right\}  \nonumber \\
&&\widetilde{\mathfrak{I}}_{\pm }^{DP\,\,\,}=2J_{p\pm k}J_{p}  \nonumber \\
&&\Re \left\{ e^{-i\left( \psi \mp \varphi \right) }R_{p\pm k}\mathbf{G}%
_{p}^{OUT\dagger }\cdot \tilde{\mathbf{G}}_{p\pm k}\cdot \mathbf{\Phi }\cdot %
\mathfrak{X}\cdot \mathbf{G}_{p\pm k}\right\}  \nonumber \\
&&\delta \widetilde{\mathbf{I}}_{\pm }^{DP\,\,DEF}=2J_{p\pm k}J_{p} 
\nonumber \\
&&\Re \left\{ e^{-i\left( \psi \mp \varphi \right) }R_{p\pm k}\mathbf{G}%
_{p}^{OUT\dagger }\cdot \tilde{\mathbf{G}}_{p\pm k}\cdot \mathbf{\Phi }\cdot
\delta \tilde{\mathbf{\varsigma}}_{1,cav}^{DEF}\cdot \mathbf{G}_{p\pm
k}\right\}  \nonumber \\
&&\tilde{X}_{\pm }^{DP\;SN}=2J_{p}\Re \left\{ \mathbf{v}_{1}^{\dagger }\cdot 
\mathbf{G}_{p}^{OUT\dagger }\cdot \delta \mathbf{\tilde{a}}_{p\pm
k}^{OUT\;SN}\right\}  \label{DP matrices}
\end{eqnarray}

\section{\label{sec:Appendix-C}Small misalignment and mismatch}

Assuming $R=1$ and $p=0$ the ponderomotive force and torques read 
\begin{eqnarray}
\bar{F}_{0} &=&\left| G_{0\left( 00,00\right) }\right| ^{2}  \nonumber \\
\bar{T}_{0,Jq} &=&2\mbox{Re}\left\{ G_{0\left( 00,00\right) }^{\ast
}G_{0\left( 10,10\right) }v_{Jq}\right\}   \nonumber
\end{eqnarray}
while the stiffness vectors reduce to 
\begin{eqnarray}
\tilde{\mathfrak{F}}_{J} &=&\tilde{\mathfrak{F}}^{\left( 0\right) }+2%
\mbox{Re}\left\{ \tilde{\mathfrak{F}}_{J}^{\left( 1\right) }\right\}  
\nonumber \\
\tilde{\mathfrak{T}}_{Jq} &=&\tilde{\mathfrak{T}}^{\left( 0\right) }+2%
\mbox{Re}\left\{ \tilde{\mathfrak{T}}_{J}^{\left( 1\right) }\right\} 
\label{stiffness small misalignments}
\end{eqnarray}
with 
\begin{eqnarray}
\tilde{\mathfrak{F}}^{\left( 0\right) } &=&\left( \tilde{F}_{\psi
},0,0,0,0\right)   \nonumber \\
\tilde{\mathfrak{T}}^{\left( 0\right) } &=&\left( 0,\tilde{T}_{X},\tilde{T}%
_{X},\tilde{T}_{Y},\tilde{T}_{Y}\right)   \nonumber \\
\tilde{\mathfrak{F}}_{J}^{\left( 1\right) } &=&\left( 0,\tilde{F}_{X}v_{Jy},%
\tilde{F}_{X}v_{Jz},\tilde{F}_{Y}v_{Jy},\tilde{F}_{Y}v_{Jz}\right)  
\nonumber \\
\tilde{\mathfrak{T}}_{J}^{\left( 1\right) } &=&\left( \tilde{T}_{\psi
}\left( v_{Jy}+v_{Jz}\right) ,0,0,0,0\right)   \label{stiffness vectors}
\end{eqnarray}
where 
\begin{eqnarray}
\tilde{F}_{\psi } &=&2\Im \left\{ e^{-i\psi -i2\phi _{G}}\tilde{G}_{0\left(
00,00\right) }\left| G_{0\left( 00,00\right) }\right| ^{2}\right\}  
\nonumber \\
\tilde{T}_{X} &=&2\Im \left\{ e^{-i\psi -i4\phi _{G}}\tilde{G}_{0\left(
10,10\right) }\left| G_{0\left( 00,00\right) }\right| ^{2}\right\}  
\nonumber
\end{eqnarray}
Similar expression holds for $\tilde{F}_{X}$ and $\tilde{T}_{\psi }$ with $%
\left| G_{0\left( 00,00\right) }\right| ^{2}$ replaced by $G_{0\left(
00,00\right) }^{\ast }G_{0\left( 10,10\right) }$ while $\tilde{F}_{Y}$ is
similar to $\tilde{F}_{X}$ with $\Im $ replaced by $\Re .$

Analogously, for the Drever-Pound error signal 
\begin{eqnarray}
\bar{s}^{DP} &=&\bar{s}^{\left( 0\right) }  \nonumber \\
\tilde{\mathfrak{s}}^{DP} &=&\tilde{\mathfrak{s}}^{\left( 0\right) }+2%
\mbox{Re}\left\{ \tilde{\mathfrak{s}}^{\left( 1\right) }\right\}   \nonumber
\\
\delta \tilde{s}_{cav}^{DP\;DEF} &=&\delta \tilde{s}_{cav}^{\left( 0\right)
\;DEF}+2\mbox{Re}\left\{ \delta \tilde{\mathfrak{s}}_{cav}^{\left( 1\right)
\;DEF}\right\}   \label{D-P expansion}
\end{eqnarray}
with 
\begin{eqnarray}
\bar{s}^{\left( 0\right) } &=&\left( \bar{s}_{\psi },0,0,0,0\right)
\label{D-P expansion coefficients} \\
\tilde{\mathfrak{s}}^{\left( 0\right) } &=&\left( \tilde{s}_{\psi
},0,0,0,0\right)   \nonumber \\
\tilde{\mathfrak{s}}^{\left( 1\right) } &=&\left( 0,\tilde{s}_{X}v_{1y},%
\tilde{s}_{X}v_{1z},\tilde{s}_{Y}v_{1y},\tilde{s}_{Y}v_{1z}\right)  
\nonumber \\
\delta \tilde{s}^{\left( 0\right) \;DEF} &=&\tilde{s}_{\psi }\left(
e^{i\omega \tau }\varsigma _{1s\left( 00,00\right) }\delta \tilde{\varsigma}%
_{1s}+e^{i\omega \tau /2}\varsigma _{2s\left( 00,00\right) }\delta \tilde{%
\varsigma}_{2s}\right)   \nonumber \\
\delta \tilde{s}^{\left( 1\right) \;DEF} &=&\tilde{s}_{X}^{\left( 1\right)
}\left( e^{i\omega \tau }\left( v_{1y}\varsigma _{1s\left( 00,01\right) }+%
\tilde{s}_{X}v_{1z}\varsigma _{1s\left( 00,10\right) }\right) \delta \tilde{%
\varsigma}_{1s}\right.   \nonumber \\
&&\left. +e^{i\omega \tau /2}\left( v_{1y}\varsigma _{2s\left( 00,01\right)
}+\tilde{s}_{X}v_{1z}\varsigma _{2s\left( 00,10\right) }\right) \delta 
\tilde{\varsigma}_{2s}\right) \nonumber
\end{eqnarray}
where 
\begin{eqnarray}
\bar{s}_{\psi } &=&J_{p+k}J_{p}2\Im \left\{ e^{i\varphi }G_{p+k\left(
00,00\right) }^{OUT\;\ast }G_{p\left( 00,00\right) }^{OUT}\right\}  
\nonumber \\
\tilde{s}_{\psi } &=&2J_{p+k}J_{p} %
\Re \left\{ e^{-i\psi -i2\phi _{G}} \right.
\nonumber \\
&&\left. \left( e^{i\varphi }R_{p+k}G_{p\left(
00,00\right) }^{OUT\;\ast }\tilde{G}_{p+k\left( 00,00\right)
}^{OUT}G_{p+k\left( 00,00\right) }^{OUT}\right. \right.   \nonumber \\
&&\left. \left. -e^{-i\varphi }R_{p}G_{p+k\left( 00,00\right) }^{OUT\;\ast }%
\tilde{G}_{p\left( 00,00\right) }^{OUT}G_{p\left( 00,00\right)
}^{OUT}\right) \right\}   \label{DP psi static & dynamic}
\end{eqnarray}
$\tilde{s}_{X}$ is similar to $\tilde{s}_{\psi }$ with $G_{p+k\left(
00,00\right) }^{OUT},G_{p\left( 00,00\right) }^{OUT}$ replaced by $%
G_{p+k\left( 10,10\right) }^{OUT},G_{p\left( 10,10\right) }^{OUT}$, while $%
\tilde{s}_{Y}$ is similar to $\tilde{s}_{X}$ with $\Re $ replaced by $\Im .$

Finally the shot noise contribution reads 
\[
\tilde{X}^{DP\;SN}=\tilde{X}^{DP\;SN\;\left( 0\right) }+\tilde{X}%
^{DP\;SN\;\left( 1\right) } 
\]
where 
\begin{eqnarray*}
\tilde{X}^{DP\;SN\;\left( 0\right) } &=&2J_{p}\Re \left\{ G_{p\left(
00,00\right) }^{OUT\ast }\left( \delta \tilde{a}_{p+k\left( 00\right)
}^{SN}-\delta \tilde{a}_{p-k\left( 00\right) }^{SN}\right) \right\} \\
\tilde{X}^{DP\;SN\;\left( 1\right) } &=&2\Re \left\{ G_{p\left( 10,10\right)
}^{OUT\ast }\left( \left( \delta \tilde{a}_{p+k\left( 10\right)
}^{SN}-\delta \tilde{a}_{p-k\left( 10\right) }^{SN}\right) v_{1y}^{\ast
}\right. \right. \\
&&\left. \left. +\left( \delta \tilde{a}_{p+k\left( 01\right) }^{SN}-\delta 
\tilde{a}_{p-k\left( 01\right) }^{SN}\right) v_{1z}^{\ast }\right) \right\}
\end{eqnarray*}

\section{\label{sec:Appendix-D}Thermal and shot-noise sources}

\begin{eqnarray}
\tilde{\mathcal{X}}_{J\psi }^{TH/SN} &=&\tilde{\chi}_{J\psi }\tilde{X}%
_{J\psi }^{TH/SN}+\tilde{\chi}_{J\psi \theta z}\tilde{X}_{J\theta z}^{TH/SN}
\nonumber \\
&&+\delta _{1J}H^{DP}\mathcal{E}^{2}\tilde{s}^{DP\;TH/SN}  \nonumber \\
\tilde{\mathcal{X}}_{Jq}^{TH} &=&\tilde{\chi}_{Jq}\tilde{X}_{Jq}^{TH} 
\nonumber \\
\tilde{\mathcal{X}}_{J\theta z}^{TH/SN} &=&\tilde{\chi}_{J\theta z\psi }%
\tilde{X}_{J\psi }^{TH/SN}+\tilde{\chi}_{J\theta z}\tilde{X}_{J\theta
z}^{TH/SN}  \nonumber \\
&&+\delta _{1J}H_{z}^{QD}\mathcal{E}^{2}\tilde{s}_{z}^{QD\;TH/SN}  \nonumber
\\
\tilde{\mathcal{X}}_{J\theta y}^{TH} &=&\tilde{\chi}_{J\theta y}\tilde{X}%
_{J\theta y}^{TH}+\delta _{1J}H_{y}^{QD}\mathcal{E}^{2}\tilde{s}_{y}^{QD\;TH}
\label{nose sources}
\end{eqnarray}
while for the mirror modes 
\[
\tilde{\mathcal{X}}_{Js}^{DEF\;TH/SN}=\chi _{Js}\tilde{X}_{Js}^{DEF\;TH/SN} 
\]

\section{\label{sec:Appendix-E}Thermal noise correlations}

The correlations $\overline{\tilde{X}^{TH}\left( \varpi \right) \tilde{X}%
^{TH}\left( \varpi ^{\prime }\right) }=C^{XX\;TH}\delta \left( \varpi
+\varpi ^{\prime }\right) ,\ldots $ of the thermal sources (\ref{thermal
noise}) and (\ref{thermal noise Y}) for the Diosi master equation (see \cite
{Jacobs}) are given by 
\begin{eqnarray}
\tilde{C}^{XX\;TH} &=&\frac{4k_{B}T}{\hbar \varpi _{J}}+\frac{\left| \varpi
+i\gamma _{J}\right| ^{2}}{\varpi _{J}^{2}}\frac{\hbar \varpi _{J}}{3k_{B}T}
\label{thermal correlations} \\
&&+2\frac{\varpi }{\varpi _{J}}  \nonumber \\
\tilde{C}^{YY\;TH} &=&\frac{4k_{B}T}{\hbar \varpi _{J}}+\frac{\varpi _{J}^{2}%
}{\varpi ^{2}}\frac{\hbar \varpi _{J}}{3k_{B}T}+2\frac{\varpi _{J}}{\varpi }
\nonumber \\
\tilde{C}^{XY\;TH} &=&\tilde{C}^{YX\;TH\ast }\left( \varpi \right) =\frac{%
4k_{B}T}{\hbar \varpi _{J}}  \nonumber \\
&&+\frac{\varpi +i\gamma _{J\lambda \hat{\jmath}}}{\varpi }\frac{\hbar
\varpi _{J}}{3k_{B}T}+\frac{\varpi _{J\lambda \hat{\jmath}}}{\varpi }+\frac{%
\varpi +i\gamma _{J\lambda \hat{\jmath}}}{\varpi _{J\lambda \hat{\jmath}}}
\nonumber
\end{eqnarray}
while for the master equation of Ref. \cite{Giovannetti} 
\begin{eqnarray*}
\tilde{C}^{XX\;TH} &=&\tilde{C}_{GV}^{YY\;TH}=\tilde{C}_{GV}^{XY\;TH} \\
&=&2\frac{\varpi }{\varpi _{J}}\left( 1+\coth \left( \frac{\hbar \varpi }{%
2K_{B}T}\right) \right) 
\end{eqnarray*}
On the other hand the commutators coincide 
\begin{equation}
\left. \overline{\left[ \tilde{X}^{DEF\;TH},\tilde{X}/\tilde{Y}^{DEF\;TH}%
\right] }\right| =4\frac{\varpi }{\varpi _{J}}  \label{thermal commutators}
\end{equation}

\section{\label{sec:Appendix-F}Shot noise correlations}

The Fourier transforms of the shot noise force and torque (\ref{suspension
shot noise}) are characterized by the correlations 
\begin{eqnarray*}
&&\overline{\tilde{X}_{Ji}^{SN}\left( \varpi \right) \tilde{X}_{J^{\prime
}i^{\prime }}^{SN}\left( \varpi ^{\prime }\right) } \\
&=&\left( 1+\frac{t_{2}^{2}}{t_{1}^{2}}\right) \tilde{C}_{JiJ^{\prime
}i^{\prime }}^{SN}\delta \left( \varpi +\varpi ^{\prime }\right)
\end{eqnarray*}
where 
\[
\tilde{C}_{JiJ^{\prime }i^{\prime }}^{SN}=\mathbf{v}_{J}^{\dagger }\mathbf{%
\cdot} \tilde{\mathbf{C}}_{ii^{\prime }}^{SN}\cdot \mathbf{v}_{J^{\prime }} 
\]
with 
\[
\tilde{\mathbf{C}}_{ii^{\prime }}^{SN}=J_{p}^{2}\mathbf{G}_{p}^{\ast }\cdot 
\mathbf{X}_{i}\cdot \left| \tilde{\mathbf{G}}_{p}\right| ^{2}\cdot \mathbf{X}%
_{i^{\prime }}\cdot \mathbf{G}_{p} 
\]
and $i,i^{\prime }=1,2,3.$ In particular, $\tilde{\mathbf{C}}_{JiJ^{\prime
}i^{\prime }}^{SN}=\tilde{\mathbf{C}}_{J^{\prime }i^{\prime
}Ji}^{SN\;\ddagger }$.

Analogously for $\tilde{X}_{Js}^{DEF\;SN}$ (see (\ref{mirror modes force})), 
\begin{eqnarray*}
&&\overline{\tilde{X}_{Js}^{DEF\;SN}\left( \varpi \right) \tilde{X}%
_{J^{\prime }s^{\prime }}^{DEF\;SN}\left( \varpi ^{\prime }\right) } \\
&=&\left( 1+\frac{t_{2}^{2}}{t_{1}^{2}}\right) \tilde{C}_{JsJ^{\prime
}s^{\prime }}^{DEF\;SN}\delta \left( \varpi +\varpi ^{\prime }\right)
\end{eqnarray*}
where 
\begin{equation}
\tilde{C}_{JsJ^{\prime }s^{\prime }}^{DEF\;SN}=\mathbf{v}_{J}^{\dagger
}\cdot \mathbf{\tilde{C}}_{JsJ^{\prime }s^{\prime }}^{DEF\;SN}\cdot \mathbf{v%
}_{J^{\prime }}  \label{shot noise correlation coefficients}
\end{equation}
with 
\begin{eqnarray}
\tilde{\mathbf{C}}_{JsJ^{\prime }s^{\prime }}^{DEF\;SN} &=&J_{p}^{2}\mathbf{G%
}_{p}^{\ast }\cdot \mathbf{\varsigma }_{Js}\cdot \left| \tilde{\mathbf{G}}%
_{p}\right| ^{2}\cdot  \nonumber \\
&&\cdot \mathbf{\varsigma }_{J^{\prime }s^{\prime }}\cdot \mathbf{G}_{p}
\label{shot noise correlation operator}
\end{eqnarray}
In addition, 
\begin{eqnarray*}
&&\overline{\left[ \tilde{X}_{Js}^{DEF\;SN}\left( \varpi \right) ,\tilde{X}%
_{J^{\prime }s^{\prime }}^{DEF\;SN}\left( \varpi ^{\prime }\right) \right] }
\\
&=&i2\left( 1+\frac{t_{2}^{2}}{t_{1}^{2}}\right) \mbox{Im}\left\{ \tilde{C}%
_{JsJ^{\prime }s^{\prime }}^{DEF\;SN}\right\} \delta \left( \varpi +\varpi
^{\prime }\right)
\end{eqnarray*}


\begin{thebibliography}{99}
\bibitem{Drever}  F.~V. Kowalski, J. Hough, G. M. Ford, A. J.~Munley,
R.~W.~Drever, J. L.~Hall and H.~Ward. \emph{Appl. Phys. B}, 31 97, (1983);

\bibitem{La Penna}  P. La Penna, A. Di Virgilio, M. Fiorentino, A. Porzio,
S. Solimeno, \emph{Opt. Commun.}, 162, 267 (1999); E. D'Ambrosio, \emph{%
Phys. Rev. D} 67, 102004 (2003);

\bibitem{Bernardini}  Bernardini et al. \emph{Phys.Lett. A}, 243:187--194,
(1998); G. Cella, V.S. Chickarmane, A. Di Virgilio, A. Gaddi, A. Vicer\'{e} 
\emph{Phys. Letters A,} 266, 1 (2000); L. Bracci et al., \emph{Class. Quant.
Grav.}, 19:1675-1682 (2002); A. Di Virgilio et al., \emph{Phys. Lett. A},
316:1-9 (2003); \_\_\_\_, \emph{Phys. Lett. A}, 318:199-204 (2003);
\_\_\_\_, \emph{Phys. Lett. A}, 322:1-9 (2004); \_\_\_\_, \emph{Class.
Quant. Grav.}, 21:S1099-S1106 (2004); \_\_\_\_, ''Considerations on
collected data with the Low Frequency Facility Experiment'', \emph{\
accepted for publication in J. Phys. Conf. Series}, (2005);

\bibitem{Fritschel}  P. Fritschel, N. Mavalvala, D. Shoemaker, D. Sigg, M.
Zucker, and G. Gonzales, \emph{Appl. Opt.} 37, 6734 (1998);

\bibitem{Dorsel}  A. Dorsel, J. D. McCullen, P. Meystre, E. Vignes, and H.
Walter, \emph{Phys. Rev. Letters} 51, 1550 (1983); A. Gozzini, F.
Maccarrone, F. Mango, I. Longo, and S. Barbarino, \emph{JOSAB} 2, 1841 (1985)

\bibitem{Moss}  L.~R.~Miller, G.~E.~Moss and R.~L. Forward. \emph{Appl. Opt.}%
, 10:2495--2498, (1971); Z.~Vager, A.~Abramowich and M.~Weksler. \emph{J.
Phys. E}, 19:182--188, (1986); J.-P. Richard, \emph{Phys. Rev. D}, 46, 2309
(1992); H.~J. Kimble, R.~Lalezari, G.~Rempe, R. J.~Thompson, \emph{Opt. Lett.%
} 17, 363 (1992).; N.~Mio and K.~Tsubono. \emph{Appl. Opt.}, 34, 186 (1995).

\bibitem{web}  Overviews of the different studies can be found on the
different projects web sites: http://www.ego-gw.it;
http://www.ligo.caltech.edu; http://www.geo600.uni--hannover.de;
http://tamago.mtk.nao.ac.jp;

\bibitem{Law}  C. K. Law, \ \emph{Phys. Rev. A} 51:2537 (1995);

\bibitem{Pace}  A. F. Pace, M. J. Collett, and D. F. Walls, \emph{Phys. Rev.}
A 47:3173 (1993)

\bibitem{Deruelle}  N. Deruelle and P. Tourrenc in \emph{Gravitation,
Geometry and Relativistic Physics,} Springer-Verlag, Berlin (1984); P.
Tourrenc and N. Deruelle \emph{Ann. Phys. (Paris)} 10, 241 (1985); J. M.
Aguirregabiria and L. Bel \emph{Phys. Rev. A} 36, 3768 (1987); L. Bel, J. L.
Boulanger and N. Deruelle \emph{Phys. Rev. A} 37, 2563 (1988); B. Meers and
N. MacDonald \emph{Phys. Rev. A} 40, 3754 (1989); S. Solimeno, F. Barone, C.
de Lisio, L. Di Fiore, L. Milano and G. Russo, \emph{Phys. Rev. A} 43, 6227
(1991);

\bibitem{Gardiner}  C.W. Gardiner, \emph{"Handbook of Stochastic Methods"},
Springer, Berlin (1985); A. O. Caldeira and A. J. Leggett, \emph{Physica}, A
121, 587 (1983); \_\_\_ \emph{Phys. Rev. A}, 31, 1059 (1985); W. G. Unruh
and W. H. Zurek, \emph{Phys. Rev. }D, 140, 1071 (1989);

\bibitem{Jacobs}  K. Jacobs, I. Tittonen, H. M. Wiseman and S. Schiller, 
\emph{Phys. Rev. A}, 60, 538 (1999); I. Tittonen, G. Breitenbach, T.
Kalkbrenner, T. Muller, R. Conradt, S. Schiller, E. Steinsland, N. Blanc,
and N. F. de Rooij, \emph{Phys. Rev.} A 59, 1038 (1999);

\bibitem{Munro}  W. J. Munro and C.W. Gardiner,\emph{\ Phys. Rev. }A, 53,
2633 (1996); \ S. Guntzmann and F. Haake, \emph{Z. Phys }B 101, 263 (1996);

\bibitem{Giovannetti}  V. Giovannetti and D. Vitali, \emph{Phys. Rev. A},
63, 23812 (2001);

\bibitem{Barchielli}  A. Barchielli, Phys. Rev. A, 34, 1642 (1986);

\bibitem{Collet}  M. J. Collet and C. W. Gardiner, \emph{Phys. Rev. A} 30,
1386 (1984); C. W. Gardiner and M. J. Collett \emph{Phys. Rev. A} 31, 3761
(1985);

\bibitem{Braginsky&Manukin}  V. Braginsky and A. Manukin,
Sov.~Phys.~JETP-USSR 25, 653-655 (1967); V. Braginsky and A. Manukin, 
\textit{\ Measurement of weak forces in Physics Experiments}, (University of
Chicago Press, 1977), pp. 25-39.

\bibitem{Sidles}  J. Sidles and D. Sigg \emph{LIGO document no. LIGO-P03005}

\bibitem{Braginsky}  V. B. Braginsky, S. E. Strigin and S. P. Vyatchanin 
\emph{Phys. Lett. A} 287, 331 (2001);\_\_\_\_ 293, 228 (2002); \_\_\_ 305,
111 (2002) W. Kells, and E. D'Ambrosio \emph{Phys. Lett. A} 299, 326 (2002);
S. W: Chedddiwy, Chunnong Zhao, Li Ju and D. G. Blair \emph{Class. Quantum
Grav.} 21, 1253 (2004)

\bibitem{spring}  V. B. Braginsky and F. Ya. Khalili, Phys. Lett. A 257, 241
(1999); Benjamin S. Sheard, et al. Phys. Rev. A, 69, 051801 (2004); H.
Rokhsari et al., Opt. Expr., 13, 5293 (2005); Thomas Corbitt et al.,
LIGO--P050045--00--R (http://www.ligo.org/pdf\_public/P050045.pdf); A. Di
Virgilio, et al. "Evidence of an optical spring" \emph{in preparation};

\bibitem{Mancini}  V. Giovannetti, S. Mancini and P. Tombesi, \emph{%
Europhys. Lett.} 54, 559 (2001); S. Mancini, V. Giovannetti, D. Vitali and
P. Tombesi, \emph{Phys. Rev. Lett.} 88, 120401 (2002); S. Mancini, D.
Vitali, V. Giovannetti, and P. Tombesi, \emph{Eur. Phys. J D.} 22:417 (2003);

\bibitem{Heidmann}  A. Heidmann and S. Reynaud, \emph{Phys. Rev.} A, 50,
4237 (1994); C. Fabre, M. Pinard, S. Bourzeix, A. Heidmann, E. Giacobino,
and S. Reynaud, \emph{Phys. Rev.} A, 49, 1337 (1994); M. Pinard, C. Fabre,
and A. Heidmann, \emph{Phys. Rev.} A, 51, 2443 (1995); P.F. Cohadon, A.
Heidmann, and M. Pinard, \emph{Phys. Rev. Letters} 83, 3174 (1999); Y.
Hadjar, P. F. Cohadon, C. G. Aminoff, M. Pinard, and A. Heidmann, \emph{%
Europhys. Letters} 47, 545 (1999); M. Pinard, P. F. Cohadon, T. Briant, and
A. Heidmann, \emph{Phys. Rev} A 63, 013808 (2000); J. M. Courty, A.
Heidmann, and M. Pinard,\emph{Phys. Rev. Letters,} Feb (2003); T. Corbitt,
K. Goda, N. Mavalvala, E. Mikhalov, D. Ottaway, S. Whitcomb, Y. Chen,
Caltech Seminar, March (2004);

\bibitem{Saulson}  P.~R. Saulson. \emph{Phys. Rev. D}, 2437 (1990); G. I.
Gonzalez and P. R. Saulson, \emph{J Acoustic Soc. Am.} 96, 207 (1994);
Gillespie, A. and Raab, R., \emph{Phys. Rev. D} 52, 577 (1995); F. Bondue
and J.-Y. Vinet, \emph{Phys. Lett. A} 198, 74 (1995); F. Bondue, P. Hello
and J-Y Vinet, \emph{Phys. Lett. A} 246, 227 (1998);

\bibitem{Hadjar}  Y. Hadjar \textquotedblright High sensitivity optical
measurement of mechanical Brownian motion of interferometric detector of
Virgo gravitational wave PhD, LAL Orsay (1999)

\bibitem{Numata}  K. Numata, M. Ando, K. Yamamoto, S. Otsuka and K. Tsubono, 
\emph{Phys. Rev. Lett.} 91, 260602 (2003)

\bibitem{Levin}  Yu. Levin, \emph{Phys. Rev. D} 57, 659 (1998);

\bibitem{Vicere}  A. Vicer\`{e} Proc. Int. Summer School on
Exp. Phys. of Gravitational Waves, World Scientific
Publishing Co. Pte. Ltd., Singapore, (2000).

\end{thebibliography}
\end{document}